\documentclass[preprint,aps]{revtex4-1}
\usepackage{graphicx}
\usepackage{hyperref}
\usepackage{epstopdf}
\usepackage{slashed}
\usepackage{rotating}
\usepackage{float}
\newcommand{\be}{\begin{equation}}
\newcommand{\ee}{\end{equation}}
\newcommand{\bea}{\begin{eqnarray}}
\newcommand{\eea}{\end{eqnarray}}
\newcommand{\ba}{\begin{array}}
\newcommand{\ea}{\end{array}}

\begin{document}
\title{Nucleon structure functions and longitudinal spin asymmetries in the chiral quark constituent model}
\author{Harleen Dahiya}
\affiliation{Department of Physics,\\ Dr. B.R. Ambedkar National
Institute of Technology,\\ Jalandhar, 144011, India}
\author{Monika Randhawa}
\affiliation{University Institute of Engineering and Technology,
Panjab University, Chandigarh, 160014, India}
\begin{abstract}

We have analysed the phenomenological dependence of the spin independent ($F_1^{p,n}$ and $F_2^{p,n}$) and the spin dependent ($g_1^{p,n}$) structure functions of the nucleon on the the Bjorken
scaling variable $x$ using the unpolarized distribution functions of the quarks $q(x)$ and the polarized distribution functions of the quarks $\Delta q(x)$ respectively. The chiral constituent quark model ($\chi$CQM), which is known to provide a satisfactory explanation of the proton
spin crisis and related issues in the nonperturbative regime, has been used to compute explicitly the valence and sea quark flavor distribution functions of $p$ and $n$. In light of the improved precision of the world data, the $p$ and $n$ longitudinal spin asymmetries ($A_1^p(x)$ and $A_1^n(x)$) have been calculated. The implication of the presence of the sea quarks has been discussed for ratio of polarized to unpolarized quark distribution functions for up and down quarks in the $p$ and $n$ $\frac{\Delta u^p(x)}{u^p(x)}$, $\frac{\Delta d^p(x)}{d^p(x)}$, $\frac{\Delta u^n(x)}{u^n(x)}$, and $\frac{\Delta d^n(x)}{d^n(x)}$. The ratio of the $n$ and $p$ structure functions $R^{np}(x)=\frac{F_2^n(x)}{F_2^p(x)}$ has also been presented. The results have been compared with the recent available experimental observations. The results on the spin sum rule have also been included and compared with data and other recent approaches.

\end{abstract}

\maketitle
\section{Introduction}
Several interesting studies have been carried out to understand the internal structure of the nucleon ever since  the  deep
inelastic scattering (DIS) experiments revealed that the quarks are point-like constituents \cite{point-like}. These  point-like constituents were identified as the valence or constituent quarks
with spin-$\frac{1}{2}$ in the naive quark model (NQM)
\cite{dgg,Isgur,yaouanc,mgupta}. Surprisingly, the measurements of polarized structure functions of proton in DIS experiments \cite{emc,smc,adams,hermes_spin} showed that the total spin carried by the constituent quarks was very small (only about 30\%) leading to the ``proton spin crisis''; see Ref. \cite{rev_spin} for a recent review.
The polarized deep inelastic lepton-nucleon scattering is an
useful probe of the spin structure of the nucleon and the measurements with proton, deuteron, and helium-3 targets
have determined the unpolarized and polarized structure functions of the nucleon through the measurement of the longitudinal spin asymmetries with the target spin being parallel and antiparallel to the longitudinally polarized beam \cite{smc,a1np}.

The data on the asymmetry of the nucleons as well as the ratio of neutron and proton unpolarized structure functions disagrees with the predictions of NQM. In addition to this, major surprise has been revealed in the famous DIS experiments by the New Muon
Collaboration (NMC) \cite{nmc}, Fermilab E866 \cite{e866}, Drell-Yan cross section ratios of the NA51 experiments \cite{baldit} and more recently by  HERMES \cite{hermes_flavor}. These experiments established the violation of Gottfried sum rule (GSR) ($\int_0^1[\bar d (x)-\bar u (x)]dx$) \cite{gsr} confirming the sea quark asymmetry of the unpolarized quarks in the case of nucleon.
Recent measurements of the electromagnetic  and weak form factors from the elastic scattering of electrons by SAMPLE at MIT-Bates \cite{sample}, G0 at JLab \cite{g0}, PVA4 at MAMI \cite{a4} and HAPPEX  at JLab \cite{happex} have given clear signals for explicit contributions of non-valence quarks in the spin structure of the nucleon. These results further confirm the nonperturbative origin of the sea quarks as the conventional perturbative production of the
quark-antiquark pairs by gluons give nearly equal numbers of antiquarks.

Even though extensive studies have been carried out in the past 40 years but it is still a big challenge to perform the calculations from the first principles of Quantum Chromodynamics (QCD). Confinement has limited our knowledge on the composition of hadrons and internal structure continues to remain a major unresolved problem in high energy spin physics. In addition to this, to have a deeper understanding of the DIS results as well as the dynamics of the constituents
of the nucleon, the ``spin sum rule'' \cite{ji} needs to be explained
\be
\frac {1}{2} =J_q+J_g=S_q+ L_{q}+J_g\,, \label{total spin}
\ee
where $S_q$ is the spin polarization contribution of
the quarks, $L_{q}$ is the orbital
angular momentum (OAM) of the quarks, $J_g$ is the
total angular momentum of the gluons.
Recently, evidence for a non zero contribution of gluon spin has been found in the polarized proton-proton collisions \cite{gluon-nonzero}.
Even though many experimental and theoretical efforts have been made to understand the contribution of OAM part, a complete understanding does not seem to have been achieved so far.

Recently, the neutrino-induced DIS experiments \cite{neudis} have emphasized that the sea quarks dominate for the values of Bjorken scaling variable $x<0.3$ and precision data have been collected only in the low and moderate $x$ regions due to experimental limitation. Further, the experiments CDHS \cite{cdhs}, CCFR \cite{ccfr1,ccfr2}, CHARMII \cite{charmii}, NOMAD \cite{nomad1,nomad2}, NuTeV \cite{nutev} and CHORUS \cite{chorus} have pointed out the need for additional refined data renewing considerable interest in the non-valence structure. In the absence of precise data above $x > 0.4$  which is a relatively clean
region to test the valence structure of the nucleon, the parametrizations are quite unconstrained. The ongoing Drell-Yan experiment at Fermilab \cite{fermilab} and a proposed experiment at J-PARC facility \cite{jparc} are working towards extending the kinematic coverage.

Considerable progress in the past few years has been made to understand the origin of the sea quarks, however, there is no consensus regarding the various mechanisms which can contribute to it.  The broader question of non-valence quark contribution to the unpolarized distributions of sea quarks, sea quark asymmetry, structure function has been discussed  in various models \cite{ellis-brodsky,alkofer,christov,diakonov,mesoncloud,wakamatsu,eccm,stat,bag,alwall,reya,chang-14}. One of the most successful nonperturbative approach is the chiral constituent quark model ($\chi$CQM) \cite{manohar,eichten}.  The basic idea is based on
the possibility that chiral symmetry breaking takes place at a distance scale much smaller than the confinement scale.
The $\chi$CQM uses the effective interaction Lagrangian approach of the strong interactions  where the effective degrees of freedom are the
valence quarks and the internal Goldstone bosons (GBs) which are coupled to the valence quarks \cite{cheng,johan,song,hd}. The $\chi$CQM successfully explains  the spin structure of the nucleon \cite{hd}, magnetic moments of octet and decuplet baryons \cite{hdmagnetic}, semileptonic weak decay parameters \cite{nsweak}, magnetic moments of nucleon resonances and $\Lambda$ resonances \cite{nres-torres}, quadrupole moment and charge radii of octet baryons \cite{charge-radii}, etc.. On the other hand, the inclusion of Bjorken scaling variable $x$ in the distributions functions has not yet been successfully derived from first principles. Instead, they are obtained by fitting parametrizations to data. Efforts have been made in developing a  model with confining potential incorporating the $x$ dependence in the valence quarks distribution functions \cite{Isgur,yaouanc,bag,mgupta}.   The $x$ dependence in the quark distribution functions has also been derived in a physical model from simple assumptions \cite{eichten,alwall}.
In view of the above developments, it become desirable to extend the applicability of $\chi$CQM by incorporating $x$ dependence phenomenologically in the unpolarized and polarized quark distribution and nucleon structure functions whose knowledge would undoubtedly provide vital clues to the distribution of the valence and sea quarks in the kinematic range thus providing vital clues to the nonperturbative aspects of QCD.

The purpose of the present communication is to determine the unpolarized distribution functions of the quarks $q$ and the polarized distribution functions of the quarks $\Delta q$ using the chiral constituent quark model ($\chi$CQM) which successfully accounts for the quantities affected by chiral symmetry breaking. The $\chi$CQM allows us to understand the explicit contributions of the valence and the sea quarks. It would be significant to analyse the dependence of various quantities by phenomenologically incorporating the Bjorken scaling variable $x$ since $x<0.3$ is a relatively clean region to test the quark sea structure. In particular, we would like to understand in detail the spin independent structure functions $F_1^{p,n}(x)$ and $F_2^{p,n}(x)$, spin dependent  structure functions $g_1^{p,n}(x)$.  The $p$ and $n$ longitudinal spin asymmetries $A_1^p(x)$ and $A_1^n(x)$ come from the difference in cross sections in scattering of a polarized lepton from a polarized proton where the leptons are scattered with the same and unlike helicity as that of the proton. Further, it would be interesting to extend the calculations to compute the ratio of polarized to unpolarized quark distribution functions for up and down quarks in the $p$ and $n$ $\frac{\Delta u^p(x)}{u^p(x)}$, $\frac{\Delta d^p(x)}{d^p(x)}$, $\frac{\Delta u^n(x)}{u^n(x)}$, and $\frac{\Delta d^n(x)}{d^n(x)}$. The implications of the presence of the sea quarks can also discussed for the ratio of the $n$ and $p$ spin independent structure functions $R^{np}(x)=\frac{F_2^n(x)}{F_2^p(x)}$. The role of valence and sea quarks and their orbital angular momentum can be discussed in the context of spin sum rule. The results can be compared with the recent available approaches and also provide important constraints on the future experiments to describe the role of non-valence degrees of freedom.

\section{Unpolarized and polarized distribution functions of quarks \label{quark-st-fns}}
The unpolarized distribution function of the quark (antiquark) $q_i(x)$ ($\bar{q_i}(x)$) is described as the probability of the $i^{th}$ quark (antiquark) carrying a fraction $x$ of the nucleon's momentum. It can be calculated from the scalar matrix element of the nucleon using the operator ${q \bar q}$ measuring the sum of the quark and antiquark numbers as
\be \langle N|{q \bar q}|N \rangle, \label{bnb}
\ee
where $|N\rangle$ is the nucleon wavefunction. The operator ${q \bar q}$ is defined in terms of the number $n_{q({\bar q})}$ of $q({\bar q})$ quarks with electric charge $e_q(e_{\bar q})$. We have
\bea {q \bar q}=
\sum_{q=u,d,s} (n_q q + n_{\bar q }{\bar q})&=& n_{u}u + n_{{\bar
u}}{\bar u} + n_{d}d + n_{{\bar d}}{\bar d} + n_{s}s + n_{{\bar
s}}{\bar s} \,. \label{number1} \eea

The polarized distribution function of the $i^{th}$ quark $\Delta q_i(x)$ is defined as
\be
\Delta q_i(x)= q^{\uparrow}_i(x)- q^{\downarrow}_i(x), \label{deltaq}
\ee
 where $q^{\uparrow}_i(x)$ $(q^{\downarrow}_i(x))$ is the probability that the $i^{th}$ quark spin is aligned parallel or antiparallel to the nucleon spin.
The polarized distribution function of the quarks can be calculated from the axial vector matrix element of the nucleon using the operator $q^{\uparrow} q^{\downarrow}$ measuring the sum of the quark with spin up and down as
\be
\langle N|q^{\uparrow}q^{\downarrow}|N \rangle\,. \label{BNB}
\ee
Here  ${\cal N}=q^{\uparrow}q^{\downarrow}$ is the number
operator defined in terms of the number $n_{q^{\uparrow}(q^{\downarrow})}$ of $q^{\uparrow}({q^{\downarrow}})$ quarks. We have
\be
q^{\uparrow}q^{\downarrow}=\sum_{q=u,d,s} (n_{q^{\uparrow}}q^{\uparrow} + n_{q^{\downarrow}}q^{\downarrow})=n_{u^{\uparrow}}u^{\uparrow} + n_{u^{\downarrow}}u^{\downarrow} + n_{d^{\uparrow}}d^{\uparrow} + n_{d^{\downarrow}}d^{\downarrow} +
n_{s^{\uparrow}}s^{\uparrow} + n_{s^{\downarrow}}s^{\downarrow}\,, \label{number2}
\ee
with the coefficients of the $q^{\uparrow\downarrow}$ giving the number of
$q^{\uparrow\downarrow}$ quarks.

\section{Chiral Constituent Quark Model}

The QCD Lagrangian describes the dynamics of light quarks ($u$, $d$, and $s$) and
gluons as
\be
{\cal{L}} = i \bar{\psi}_L \slashed{D}
{\psi}_L + i
\bar{\psi}_R \slashed{D} {\psi}_R  - \bar{\psi}_L M {\psi}_R - \bar{\psi}_R M {\psi}_L - \frac{1}{4}G_{\mu \nu}^{a} G^{\mu \nu}_{a} \,,
\label{lagrang1} \ee where $D^{\mu}$ is the gauge-covariant derivative, $M$ is the quark mass matrix, $\psi_L$  and $\psi_R$ are the left and right
handed quark fields respectively, and  $ G_{\mu \nu}^{a}$ is the gluonic gauge field strength tensor.
The Lagrangian in Eq. (\ref{lagrang1}) does not remain invariant under the chiral transformation $(\psi \to \gamma^5 \psi)$ as the mass
terms change sign  as $\psi_{L}
\to -\psi_{L}$ and $\psi_{R} \to \psi_{R}$. The Lagrangian will have global chiral symmetry of
the {\it SU}(3)$_L$$\times${\it SU}(3)$_R$ group if the mass terms are neglected. Around the scale of 1 GeV the chiral symmetry is believed to be spontaneously broken to ${\it SU}(3)_{L+R}$. As a consequence, there
exists a set of massless particles, referred to as the Goldstone
bosons (GBs), which are identified with the observed ($\pi$, $K$,
$\eta$ mesons). Within the region of QCD confinement scale
($\Lambda_{QCD} \simeq 0.1-0.3$ GeV) and the chiral symmetry breaking scale
$\Lambda_{\chi SB}$, the constituent quarks, the octet of GBs
($\pi$, K, $\eta$ mesons), and the {\it weakly} interacting gluons
are the appropriate degrees of freedom.

The effective interaction Lagrangian  between
GBs and quarks in the leading order can now be expressed as \be {\cal
L}_{{\rm int}} = -\frac{g_{A}}{f_{\pi}} \bar{\psi} \partial_{\mu}
\Phi \gamma^{\mu} \gamma^{5} \psi \,, \label{lagrang3} \ee where the field $\Phi$ describes the dynamics
of octet of GBs.
The QCD Lagrangian is also invariant under the axial
$U(1)$ symmetry, which would imply the existence of ninth GB. This
breaking symmetry picks the $\eta'$ as the ninth GB. The effective
Lagrangian describing interaction between quarks and a nonet of GBs,
consisting of octet and a singlet, can now be expressed as \be {\cal
L}_{{\rm int}} = g_8 { \bar \psi} \Phi {\psi} + g_1{ \bar \psi}
\frac{\eta'}{\sqrt 3}{\psi}= g_8 {\bar \psi}\left( \Phi + \zeta
\frac{\eta'}{\sqrt 3}I \right) {\psi }=g_8 {\bar \psi} \left(\Phi'
\right) {\psi} \,, \label{lagrang4} \ee where $\zeta=g_1/g_8$, $g_1$ ($g_8$)
is the coupling constant for the singlet (octet) GB and $I$ is the $3\times 3$
identity matrix.

The basic idea in the $\chi$CQM \cite{manohar} is the fluctuation process where the
GBs are emitted by a constituent quark. These GBs further splits into a $q
\bar q$ pair, for example,
\be q^{\uparrow(\downarrow)} \rightarrow {\rm GB}^0 + q^{'
\downarrow(\uparrow)} \rightarrow (q \bar q^{'})^0 +q^{'\downarrow(\uparrow)}\,, \label{basic}
\ee
where $q \bar q^{'} +q^{'}$
constitute the sea quarks \cite{cheng,johan,hd}. The GB field can be expressed in terms of the GBs and their transition probabilities as \bea
\Phi' = \left( \ba{ccc} \frac{\pi^0}{\sqrt 2}
+\beta\frac{\eta}{\sqrt 6}+\zeta\frac{\eta^{'}}{\sqrt 3} & \pi^+
  & \alpha K^+   \\
\pi^- & -\frac{\pi^0}{\sqrt 2} +\beta \frac{\eta}{\sqrt 6}
+\zeta\frac{\eta^{'}}{\sqrt 3}  &  \alpha K^o  \\
 \alpha K^-  &  \alpha \bar{K}^0  &  -\beta \frac{2\eta}{\sqrt 6}
 +\zeta\frac{\eta^{'}}{\sqrt 3} \ea \right). \eea
The transition probability of chiral
fluctuation  $u(d) \rightarrow d(u) + \pi^{+(-)}$, given in terms of the coupling constant for the octet GBs $|g_8|^2$, is defined as $a$ and is introduced by considering nondegenerate quark masses $M_s > M_{u,d}$. The probabilities of transitions of $u(d) \rightarrow s + K^{+(0)}$, $u(d,s)\rightarrow u(d,s) + \eta$, and $u(d,s) \rightarrow u(d,s) + \eta^{'}$ are given as $\alpha^2 a$, $\beta^2 a$ and $\zeta^2 a$ respectively \cite{cheng,johan}. The probability parameters $\alpha^2 a$ and $\beta^2 a$ are introduced by considering nondegenerate GB masses $M_{K},M_{\eta}> M_{\pi}$ and the probability $\zeta^2 a$ is introduced by considering  $M_{\eta^{'}} > M_{K},M_{\eta}$.

The sea quark flavor distribution functions can be calculated in $\chi$CQM by substituting for every
valence (constituent) quark \be q \to  P_q q + |\psi(q)|^2, \label{substitute}\ee
where the transition probability of no emission of
GB $P_q$ can be expressed in terms of the transition probability of the emission of a GB from
any of the $u$, $d$, and $s$ quark as follows
\be
P_q=1-P_{[q, ~GB]}, \label{probability} \ee with \be P_{[u, ~GB]}=P_{[d, ~GB]}
=\frac{a}{6}\left(9+6\alpha^2+\beta^2+2\zeta^2\right)\,,~~~~{\rm and}~~~~
P_{[s, ~GB]} = \frac{a}{3}\left(6 \alpha^2+2\beta^2+\zeta^2 \right)\,. \label{probuds} \ee
The transition
probability of the $q$  quark $|\psi(q)|^2$ calculated from the Lagrangian can be expressed as
\bea |\psi(u)|^2 &=& \frac{a}{36}\left[\left(63 +6\beta +12\zeta+4\beta \zeta +36\alpha^2+7\beta^2+16\zeta^2\right){ u} +\left(9 +6\beta+12\zeta+4\beta \zeta +\beta^2+4\zeta^2\right){ {\bar u}}\right.\nonumber\\ &+& \left.\left(45 -6\beta -12\zeta+4\beta \zeta+\beta^2+4\zeta^2\right)({d}+{ {\bar d}})+ 4\left(-2\beta \zeta +9\alpha^2+\beta^2+\zeta^2\right)({ s}+{ {\bar s}})\right]\,,\label{eqpsiu}\eea
\bea |\psi(d)|^2 &=& \frac{a}{36}\left[\left(63 +6\beta +12\zeta+4\beta \zeta +36\alpha^2+7\beta^2+16\zeta^2\right){ d} +\left(9 +6\beta+12\zeta+4\beta \zeta +\beta^2+4\zeta^2\right){ {\bar d}}\right.\nonumber\\ &+& \left.\left(45 -6\beta -12\zeta+4\beta \zeta+\beta^2+4\zeta^2\right)({u}+{ {\bar u}})+ 4\left(-2\beta \zeta +9\alpha^2+\beta^2+\zeta^2\right)({ s}+{ {\bar s}})\right]\,,\label{eqpsid}\eea
\bea |\psi(s)|^2 &=& \frac{a}{9}\left[\left(4\beta \zeta +18\alpha^2 +10\beta^2+4\zeta^2\right){ s} +\left(4\beta \zeta+4\beta^2+\zeta^2{9}\right){ {\bar s}}\right.\nonumber\\ &+& \left.\left(-2\beta \zeta +9\alpha^2+\beta^2+\zeta^2\right)({u}+{ {\bar u}}+{ d}+{ {\bar d}})\right]\,.\label{eqpsis}\eea

The spin structure of the nucleon after the inclusion of sea quarks generated
through chiral fluctuation can be calculated by substituting for
each valence (constituent) quark \be q^{\uparrow\downarrow} \rightarrow P_q q^{\uparrow\downarrow}+
|\psi(q^{\uparrow\downarrow})|^2 \,, \label{spin} \ee where  $|\psi(q^{\uparrow\downarrow})|^2$ is the probability of
transforming $q^{\uparrow\downarrow}$ quark after one interaction expressed by the
functions \bea |\psi(u^{\uparrow\downarrow})|^2 &=& \frac{a}{6}\left(3 +
\beta^2 + 2 \zeta^2 \right)u^{\downarrow\uparrow}+ a d^{\downarrow\uparrow} + a \alpha^2
s^{\downarrow\uparrow}\,, \nonumber \\ |\psi(d^{\uparrow\downarrow})|^2
&=& a u^{\downarrow\uparrow}+ \frac{a}{6} \left(3+\beta^2+2 \zeta^2 \right)d^{\downarrow\uparrow}+
a \alpha^2 s^{\downarrow\uparrow}\,, \nonumber\\ |\psi(s^{\uparrow\downarrow})|^2 &=& a \alpha^2 u^{\downarrow\uparrow} + a\alpha^2 d^{\downarrow\uparrow} +
\frac{a}{3} \left(2 \beta^2 + \zeta^2 \right)s^{\downarrow\uparrow} \,.
\label{psiup} \eea

\section{spin independent and spin dependent structure functions of the nucleon}

The nucleon structure is conventionally parameterized by the spin independent structure functions $F_1(x)$ and $F_2(x)$, and by the spin dependent structure functions $g_1(x)$ and $g_2(x)$, where  $x$ is the Bjorken scaling variable. One
useful probe of the nucleon spin structure is the longitudinal spin asymmetry
$A_1(x)$. The scattering of a polarized lepton from a polarized proton can be used to measure the spin dependent structure function $g_1$ from the difference in cross sections for leptons with the same and unlike helicity as that of the proton. The longitudinal spin asymmetries can be defined as
\be
A_1(x)=\frac{\sigma^{\uparrow\uparrow}-\sigma^{\uparrow\downarrow}}{\sigma^{\uparrow\uparrow}+\sigma^{\uparrow\downarrow}}\simeq \frac{g_1(x)}{F_2(x)}.
\ee
The spin independent  structure functions of the nucleon can be further defined in terms of the unpolarized distribution functions of the quarks defined in Sec. \ref{quark-st-fns} as follows
\bea F_1^N(x) &=& \frac{1}{2} \sum_{u,d,s} e^2_{i}(q_i(x)+\bar{q_i}(x)) \,,\nonumber \\ F_2^N(x) &=& 2xF_1^N(x) \,. \label{f12}  \eea

In the $\chi$CQM, the unpolarized distribution function of the quarks can be defined in terms of the constituent or valence as well as the sea quark distribution functions as
\be q^N(x)=q^N_{{\rm V}}(x)+q^N_{{\rm S}}(x)\,, \ee
where $q=u,d,s$.
Since the antiquark distribution functions come purely from the sea quarks  therefore we can replace the sea quark distribution functions with the antiquark distribution functions as
\be q^N(x)=q^N_{{\rm V}}(x)+\bar q^N(x). \label{totalquark}\ee
Here we have the valence quark distribution functions for $p$ and $n$ as
\bea
\int_0^1 u^p_{{\rm V}}(x)dx=2, &~~~& \int_0^1 d^p_{{\rm V}}(x)dx=1,~~~ \int_0^1 s^p_{{\rm V}}(x)dx=0\,, \nonumber \\
\int_0^1 u^n_{{\rm V}}(x)dx=1, &~~~& \int_0^1 d^n_{{\rm V}}(x)dx=2, ~~~\int_0^1 s^n_{{\rm V}}(x)dx=0\,,
\eea
and the sea quark distribution functions for $p$ and $n$ as
\bea \bar u^p= \frac{a}{12}\left(21+\beta^2+4 \zeta+4\zeta^2+\beta(2+4 \zeta)\right)\,, &~~~& \bar u^n = \frac{a}{12}\left(33+\beta^2-4 \zeta+4\zeta^2+\beta(-2+4 \zeta)\right)\,,\nonumber \\
  \bar d^p= \frac{a}{12}\left(33+\beta^2-4 \zeta+4\zeta^2+\beta(-2+4 \zeta)\right)\,, &~~~& \bar d^n = \frac{a}{12}\left(21+\beta^2+4 \zeta+4\zeta^2+\beta(2+4 \zeta)\right)\,, \nonumber \\
  \bar s^p= 3a\left(\alpha^2+\frac{1}{9}(\beta- \zeta)^2\right)\,, &~~~& \bar s^n =3a\left(\alpha^2+\frac{1}{9}(\beta- \zeta)^2\right)\,.
  \eea

There are no simple or straightforward rules which could
allow incorporation of $x$ dependence in the valence quarks and the sea quarks. For the case of unpolarized valence quark distribution function, we have incorporated the $x$ dependence phenomenologically  \cite{eichten,Isgur,yaouanc} as follows
  \bea u_{{\rm V}}^p(x)&=& 8
(1-x)^3 \cos^2 \phi+4 (1-x)^3\sin^2 \phi+ 8 \sqrt{2} x^4 (1-x)^3
\cos \phi \sin \phi\,, \label{uval} \nonumber \\ d_{{\rm V}}^p(x)&=& 4
(1-x)^3 \cos^2 \phi+2 (1-x)^3 \sin^2 \phi- 8 \sqrt{2} x^4 (1-x)^3
\cos \phi \sin \phi\,. \label{dval}
\eea
For the case of unpolarized sea quark distribution function, we have for proton \be \bar u^p(x)
=\bar u^p (1-x)^{10}\,,~~~~\bar d^p(x) =\bar d^p (1-x)^7\,, ~~~~\bar s^p(x) =\bar s^p
(1-x)^8\,, \ee
and neutron
 \be  \bar u^n(x)=\bar u^n (1-x)^{7}\,, ~~~~\bar d^n(x) =\bar d^n (1-x)^{10}\,, ~~~~\bar s^n(x) =\bar s^n
(1-x)^8\,.\ee

Using the unpolarized quark distribution
functions from Eqs. (\ref{substitute}) and (\ref{totalquark}), the structure function $F_2$ for the $p$ and $n$ Eq. (\ref{f12}) can be
expressed as
\bea  F^p_2(x) &=&\frac{4}{9} x\left(u^p_{{\rm V}}(x)+ 2\bar u^p(x)\right)
+\frac{1}{9} x\left(d^p_{{\rm V}}(x)+ 2 \bar d^p(x)+ s^p_{{\rm V}}(x)+2\bar s^p(x)\right)\,, \nonumber\\
 F^n_2(x) &=&\frac{4}{9} x\left(u^n_{{\rm V}}(x)+ 2\bar u^n(x)\right)
+\frac{1}{9} x\left(d^n_{{\rm V}}(x)+ 2 \bar d^n(x)+ s^n_{{\rm V}}(x)+2\bar s^n(x)\right)\,.
 \eea

The spin dependent  structure function of the nucleon can similarly be defined in terms of the polarized distribution function of the quarks Eq. (\ref{deltaq}) as
\be g_1^N(x) = \frac{1}{2} \sum_{u,d,s} e^2_{i}\Delta q_i(x) \,.  \ee
The polarized distribution function of the quarks can also be define in terms of polarized valence and sea quark distribution functions as
\be \Delta q^N(x)=\Delta q^N_{{\rm V}}(x)+\Delta q^N_{{\rm S}}\,. \label{totalquarksea}\ee
Here we have the polarized valence quark distribution functions for $p$ and $n$ as
\bea
\Delta u^p_{{\rm V}}=\frac{4}{3}, &~~~& \Delta d^p_{{\rm V}}=-\frac{1}{3},~~~ \Delta s^p_{{\rm V}}=0\,, \nonumber \\
\Delta u^n_{{\rm V}}=-\frac{1}{3}, &~~~& \Delta d^n_{{\rm V}}=\frac{4}{3}, ~~~\Delta s^n_{{\rm V}}=0\,, \label{deltaqval}
\eea
and the polarized sea quark distribution functions for $p$ and $n$ as
\bea
\Delta u^p_{{\rm S}}= -\frac{a}{3} (7+4 \alpha^2+
 \frac{4}{3}\beta^2 +\frac{8}{3} \zeta^2)\,, &~~~& \Delta u^n_{{\rm S}}=-\frac{a}{3} (2-\alpha^2
-\frac{1}{3}\beta^2 -\frac{2}{3} \zeta^2)\,,  \nonumber \\
\Delta d^p_{{\rm S}}=-\frac{a}{3} (2-\alpha^2
-\frac{1}{3}\beta^2 -\frac{2}{3} \zeta^2)\,, &~~~& \Delta d^n_{{\rm S}}= -\frac{a}{3} (7+4 \alpha^2+
 \frac{4}{3}\beta^2 +\frac{8}{3} \zeta^2)\,,\nonumber \\
\Delta s^p_{{\rm S}}= -a \alpha^2\,, &~~~& \Delta s^n_{{\rm S}} = -a \alpha^2\,. \label{deltaqsea}
\eea

Following Brodsky {\it et al.} \cite{brodsky}, for the polarized valence quark distribution functions of $p$ and $n$ we have parametrized
\be
\Delta u^p_{{\rm V}}(x)=\Delta u^p_{{\rm V}} (1-x)^3\,, ~~~~ \Delta d^p_{{\rm V}}(x)=\Delta d^p_{{\rm V}} (1-x)^3\,, ~~~~ \Delta s^p_{{\rm V}}(x)=\Delta s^p_{{\rm V}} (1-x)^3\,, \ee
\be
\Delta u^n_{{\rm V}}(x)=\Delta u^n_{{\rm V}} (1-x)^3\,, ~~~~ \Delta d^n_{{\rm V}}(x)=\Delta d^n_{{\rm V}} (1-x)^3\,, ~~~~ \Delta s^n_{{\rm V}}(x)=\Delta s^n_{{\rm V}} (1-x)^3\,, \ee
and for the polarized sea quark distribution functions of $p$ and $n$ we have parametrized
\be
\Delta u^p_{{\rm S}}(x)=\Delta u^p_{{\rm S}} (1-x)^5\,, ~~~~ \Delta d^p_{{\rm S}}(x)=\Delta d^p_{{\rm S}} (1-x)^5\,, ~~~~ \Delta s^p_{{\rm S}}(x)=\Delta s^p_{{\rm S}} (1-x)^5\,, \ee
\be
\Delta u^n_{{\rm S}}(x)=\Delta u^n_{{\rm S}} (1-x)^5\,, ~~~~ \Delta d^n_{{\rm S}}(x)=\Delta d^n_{{\rm S}} (1-x)^5\,, ~~~~ \Delta s^n_{{\rm S}}(x)=\Delta s^n_{{\rm S}} (1-x)^5\,. \ee

The structure function $g_1$ for $p$ and $n$ can respectively be calculated using the above equations and are expressed as
\bea  g^p_1(x) &=&\frac{4}{9} \left(\Delta u^p\right)
+\frac{1}{9} \left(\Delta d^p+ \Delta s^p\right)\,, \nonumber\\
 g^n_1(x) &=&\frac{4}{9} \left(\Delta u^n\right)
+\frac{1}{9} \left(\Delta d^n+\Delta s^n\right)\,.
 \eea

After having formulated the $x$ dependence in the valence and sea
quark distribution functions, we now consider the quantities which
are measured at different $x$ and can expressed in terms of the
above mentioned quark distribution functions.
The proton and neutron longitudinal spin asymmetries are given by
\be
A_1^p(x)=\frac{4 \Delta u^p(x)+\Delta d^p(x)}{4u^p(x)+d^p(x)}\,, ~~~~ A_1^n(x)=\frac{4 \Delta u^n(x)+ \Delta d^n(x)}{4u^n(x)+d^n(x)}\,.
\ee
These expressions can be rearranged to obtain the explicit ratio of polarized to unpolarized quark distribution functions for up and down quarks in the proton and neutron as

\bea
\frac{\Delta u^p(x)}{u^p(x)} &=& \frac{4}{15} A_1^p(x) \left( 4+\frac{d^p(x)}{u^p(x)}\right)-\frac{1}{15} A_1^n(x) \left( 1+4\frac{d^p(x)}{u^p(x)}\right)\,, \nonumber \\
\frac{\Delta d^p(x)}{d^p(x)} &=& \frac{4}{15} A_1^n(x) \left( 4+\frac{u^p(x)}{d^p(x)}\right)-\frac{1}{15} A_1^p(x) \left( 1+4\frac{u^p(x)}{d^p(x)}\right)\,,\nonumber \\
\frac{\Delta u^n(x)}{u^n(x)} &=& \frac{4}{15} A_1^p(x) \left(1+ 4\frac{d^n(x)}{u^n(x)}\right)-\frac{1}{15} A_1^n(x) \left( 4+\frac{d^p(x)}{u^p(x)}\right)\,, \nonumber \\
\frac{\Delta d^n(x)}{d^n(x)} &=& \frac{4}{15} A_1^n(x) \left(1+ 4\frac{u^n(x)}{d^n(x)}\right)-\frac{1}{15} A_1^p(x) \left( 4+\frac{u^n(x)}{d^n(x)}\right) \,.
\eea

Another important quantity where the NQM disagrees with the data significantly is the ratio of the neutron and proton structure functions
\be
R^{np}(x)=\frac{F_2^n(x)}{F_2^p(x)}\,.
\ee

\section{spin sum rule}
The various terms in the ``spin sum rule'' ($S_q+ L_{q}+J_g$) can be expressed in terms of the quantities and $\chi$CQM parameters discussed above. The spin contribution of
the quarks $S_q$ to $p$ and $n$ can be further expressed as sum of the valence and sea contributions as
\be
S^{p,n}_q= S^{p,n}_{q\rm V}+S^{p,n}_{q\rm S}\,, \label{spin_total spin}
\ee
where \bea
S^{p,n}_{q\rm V}&=&\frac{1}{2} \left (\Delta u^{p,n}_{\rm V}+\Delta d^{p,n}_{\rm V}+\Delta s^{p,n}_{\rm V} \right)\,, \nonumber \\
S^{p,n}_{q\rm S}&=&\frac{1}{2} \left (\Delta u^{p,n}_{\rm S}+\Delta d^{p,n}_{\rm S}+\Delta s^{p,n}_{\rm S} \right)\,. \eea
$S^{p,n}_{q\rm V}$ and $S^{p,n}_{q\rm S}$ for the case of $p$ and $n$ can be calculated using the polarized valence quark distribution functions and the polarized sea quark distribution functions from Eqs. (\ref{deltaqval}) and (\ref{deltaqsea}).

The total OAM carried by the quarks in the nucleon is given in terms of the transition probability of the emission of a GB from
any of the $u$, $d$, and $s$ quark $P_{[q, ~GB]}$ \cite{song-ijmpa}. We have for the case of $p$ and $n$
\bea
L_q^{p}&=& \sum_{q=u,d,s} \Delta q_{\rm V}^p P_{[q, ~GB]}\,, \nonumber \\
L_q^{n}&=& \sum_{q=u,d,s} \Delta q_{\rm V}^n P_{[q, ~GB]}\,.
\eea
In the present context, the total orbital angular momentum can be expressed in terms
of the $\chi$CQM parameters as
\be
L^{p,n}_{q}= \frac{a}{6}(9+6 \alpha^2+\beta^2+2 \zeta^2)\,.
\label{ang}
\ee
There is no direct way to calculate the contribution of gluons to the spin sum rule in the $\chi$CQM and it is already clear that gluon spin is not large enough to explain the spin problem. Therefore, we have not discussed the gluon contribution in the present work.

\section{results and discussion}
In order to study the phenomenological quantities pertaining to the valence and sea quarks distribution
functions and further compare the $\chi$CQM results with other model calculations and the available experimental data, we can study the $x$ dependence of the spin independent and spin dependent structure functions.  To this end, we first fix the $\chi$CQM parameters which provide the basis to
understand the extent to which the sea quarks contribute to the structure of the nucleon. The probabilities of fluctuations to
pions, $K$, $\eta$, $\eta^{'}$ coming in the sea quark
distribution functions are represented by $a$, $a \alpha^2$, $a \beta^2$, and $a \zeta^2$ respectively and can be obtained by taking into account strong physical considerations and carrying out a fine grained analysis using the well known experimentally measurable quantities pertaining to the spin and flavor distribution functions. The hierarchy for the probabilities, which scale as $\frac{1}{M_q^2}$, can be obtained as
\be a> a \alpha^2 \geq a \beta^2> a \zeta^2.
\ee  The mixing angle $\phi$ is fixed from the
consideration of neutron charge radius \cite{dgg}. The input
parameters and their values have been summarized in Table \ref{input}.

\begin{table}
\begin{center}
\begin{tabular}{|c|c|}      \hline
Input Parameters &Value \\ \hline
$a$ & 0.114 \\
$a \alpha^2$ & 0.023 \\
$a \beta^2$ & 0.023\\
$a\zeta^2$ & 0.002 \\
$\phi$ & $18^0$ \\  \hline
\end{tabular}
\end{center}
\caption{ Input parameters. }
\label{input}
\end{table}

\begin{figure}
\includegraphics [width=1.\textwidth] {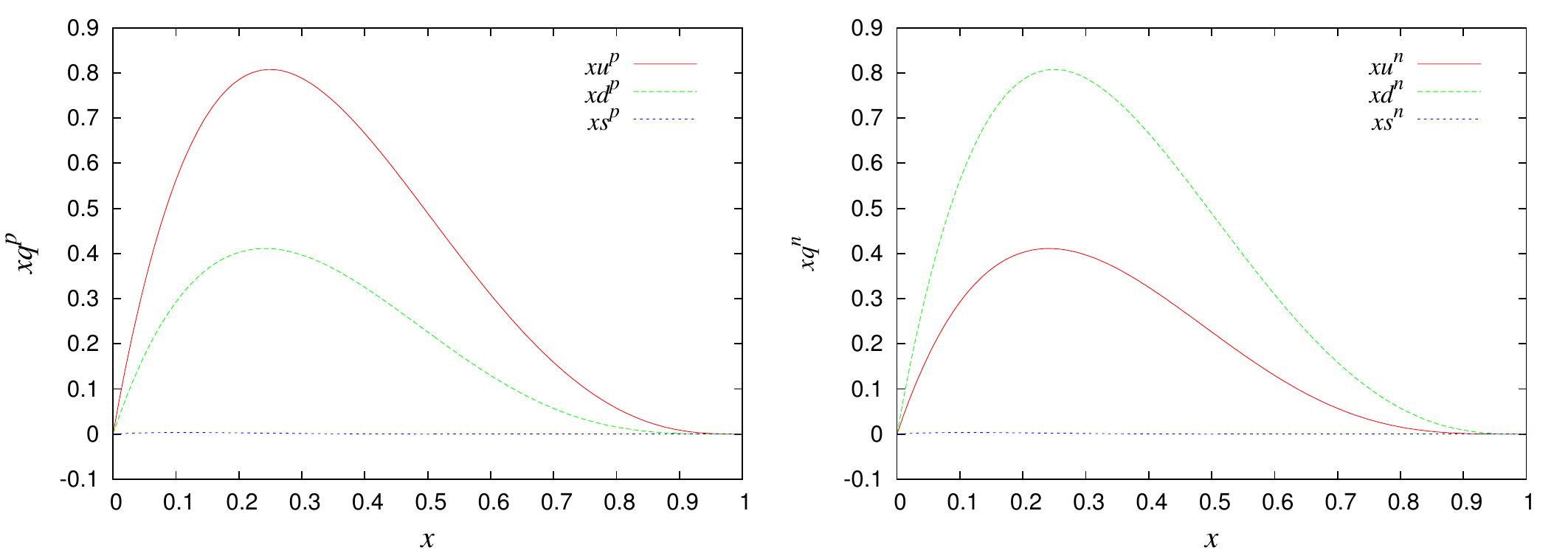}
\caption{(color online). The unpolarized quarks distribution functions for $p$:  $ x u^p(x)$, $ x d^p(x)$, $ x s^p(x)$ and $n$: $ x u^n(x)$, $ x d^n(x)$, $ x s^n(x)$ as a function of $x$.}
\label{fig1}
\end{figure}

\begin{figure}
\includegraphics [width=1.\textwidth] {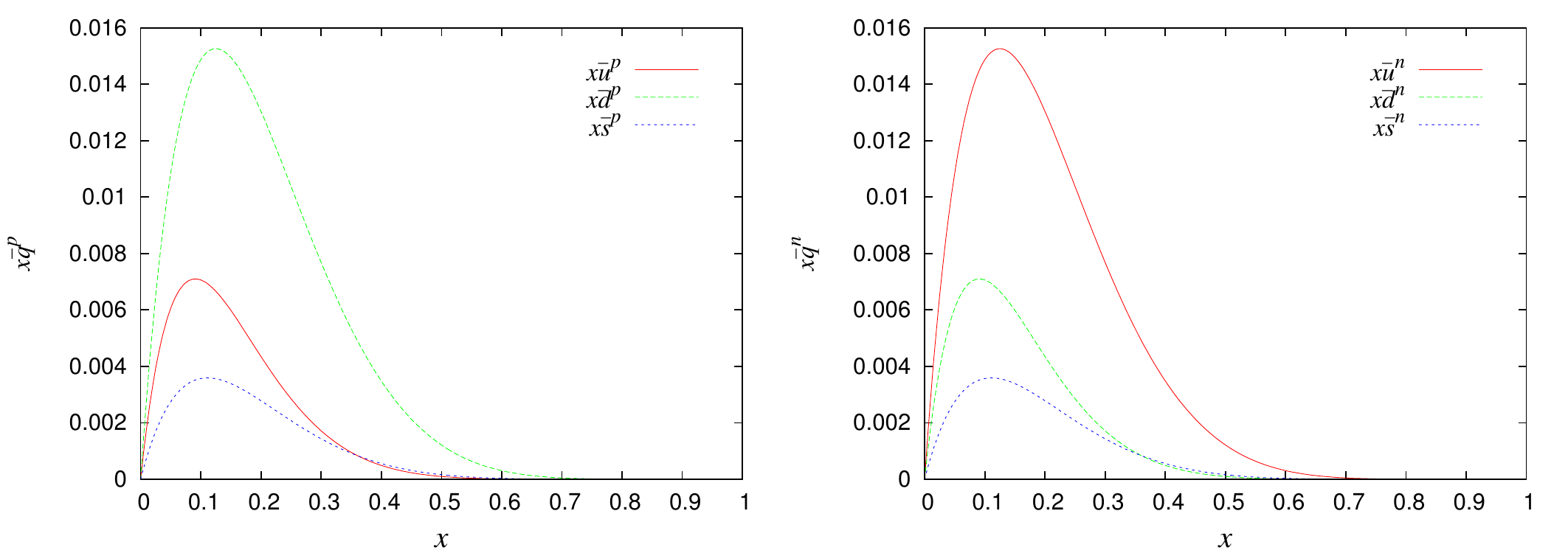}
\caption{(color online). The sea quark distribution functions for $p$:  $ x \bar u^p(x)$, $ x \bar d^p(x)$, $ x \bar s^p(x)$ and $n$: $ x \bar u^n(x)$, $ x \bar d^n(x)$, $ x \bar s^n(x)$ as a function of $x$.}
\label{fig-anti}
\end{figure}

\begin{figure}
\includegraphics [width=1.\textwidth] {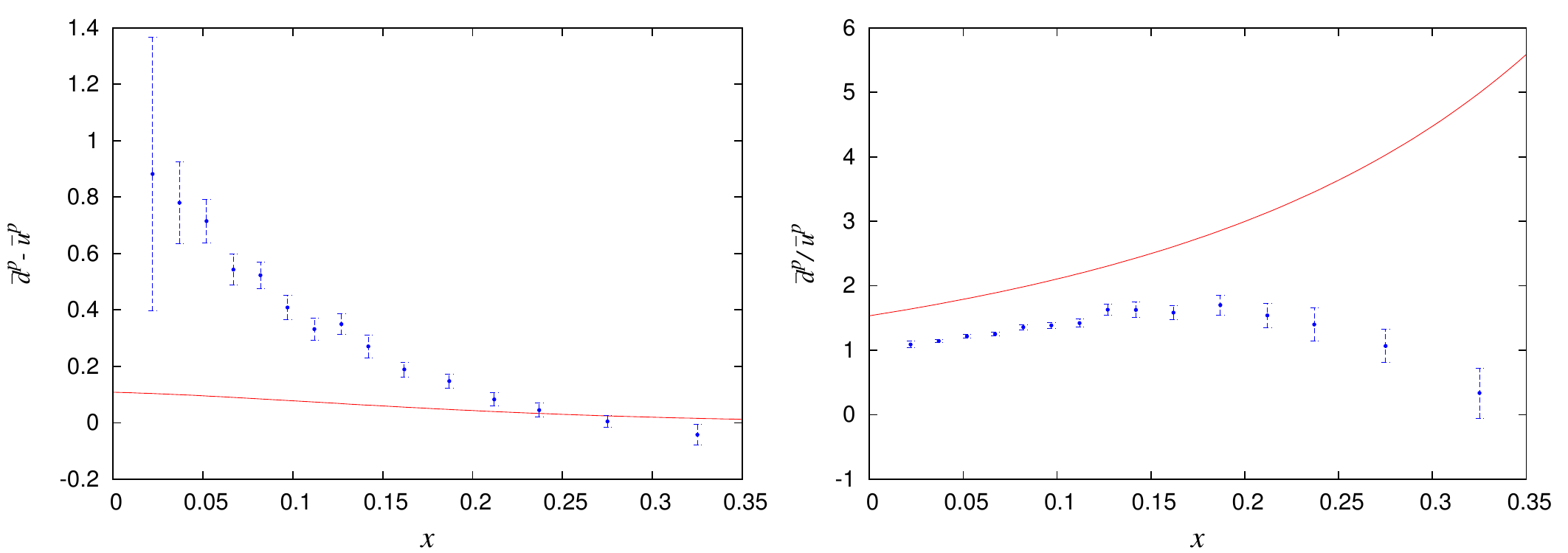}
\caption{(color online). The sea quark flavor asymmetries for the case of $p$:  $ \bar d^p(x)-\bar u^p(x)$ and $\bar u^p(x)/\bar d^p(x)$ as a function of $x$ compared with the experimental data \cite{e866}.}
\label{fig-u-d}
\end{figure}

After having incorporated $x$ dependence in the valence and the sea quark distribution functions, we now discuss the variation of all the related phenomenological quantities in the range $0\leq x \leq 1$. In Fig. \ref{fig1}, we have presented the spin independent quark distribution functions for the case $p$ ($ x u^p(x)$, $ x d^p(x)$ and $ x s^p(x)$) and $n$ ($ x u^n(x)$, $ x d^n(x)$ and $ x s^n(x)$). The valence quarks distribution functions of $p$ and $n$ vary as
\[u^p_{{\rm V}}(x)>d^p_{{\rm V}}(x)>s^p_{{\rm V}}(x),\]
\[d^n_{{\rm V}}(x)>u^n_{{\rm V}}(x)>s^n_{{\rm V}}(x).\]
On the other hand, the sea quark distribution functions vary as
\[d^p_{{\rm S}}(x)>u^p_{{\rm S}}(x)>s^p_{{\rm S}}(x),\]
\[u^n_{{\rm S}}(x)>d^n_{{\rm S}}(x)>s^n_{{\rm S}}(x).\]
It is evident from Fig. \ref{fig1} that there is $u$ quark dominance in the case of $p$ and $d$ quark dominance in the case of $n$. Since the total quark distribution functions are dominated by the valence quarks, the overall variation of the quark distribution functions is similar to the valence quark distribution functions. The variation of sea quarks distribution functions of $p$ and $n$ have been plotted in Fig. \ref{fig-anti}. Even though the variation of sea quarks if different, for example, $\bar d^p(x)$ dominates in the case of $p$ and $\bar u^n(x)$ dominates in the case of $n$, but since the probability for the fluctuation of valence quarks to sea quarks depends upon the $\chi$CQM parameter $a$ and this probability of the occurrence of sea quarks cannot be more than 10-15\%. Therefore, $u^p(x)$ dominates in the case of $p$ and $d^n(x)$ dominates in the case of $n$. This observation can also be directly related to the measurement of the Gottfried integral for the
case of nucleon which has shown a clear violation of GSR from $\frac{1}{3}$. The  quark sea asymmetry $ \int_0^1 (\bar d(x) -\bar u(x))dx$ which has been measured in the NMC and E866 experiments \cite{nmc,e866}. The NMC has reported $I^{p n}_G = \frac{1}{3} + \frac{2}{3}
\left[ \bar u^p- \bar d^p \right]=0.266 \pm 0.005$ \cite{nmc} and the E866 has reported $I^{p n}_G = 0.254 \pm 0.005$ \cite{e866}.
A flavor symmetric sea ($\bar u^N$=$\bar d^N$) would lead to $I^{p n}_G=\frac{1}{3}$.
The $\chi$CQM result for the case of nucleon ($I^{p n}_G = 0.254$) is in good
agreement with the available experimental data of E866 \cite{e866}. We have plotted some of the well
known experimentally measurable quantities, for example, $\bar d^p(x)- \bar u^p(x)$ and ${\bar d^p(x)}/{\bar u^p(x)}$ in Fig. \ref{fig-u-d} and compared them with data \cite{e866}.
It is clear from the plots that when $x$ is small $\bar d^{p}(x)-\bar u^{p}(x)$ asymmetry is large implying
the dominance of sea quarks in the low $x$ region. In fact, the
sea quarks dominate only in the region where $x$ is smaller than
0.3. At the values $x>0.3$, $\bar d-\bar u$ tends to 0 implying
that there are no sea quarks in this region.  To test the validity of the model as well as for the sake of
completeness, we can present the results of our calculations for $\bar d^p(x)- \bar u^p(x)$ and ${\bar d^p(x)}/{\bar u^p(x)}$ whose data is
available over a range of $x$ or at an average value of $x$.  We find a good overall agreement with the
data in these cases also. The data for $\bar d^p(x)- \bar u^p(x)$ is available for the
ranges $x=0-1$ and $x=0.05-0.35$ and is given as $\int_0^1{(\bar
d^p(x)-\bar u^p(x))dx}=0.118 \pm 0.012$ and $\int_{0.05}^{0.35}{(\bar
d^p(x)-\bar u^p(x))dx}=0.0803 \pm 0.011$. We find that, in our model, $\int_0^1{(\bar
d^p(x)-\bar u^p(x))dx}=0.117$ and $\int_{0.05}^{0.35}{(\bar
d^p(x)-\bar u^p(x))dx}=0.08$ in these given $x$ ranges. The valence quark distribution however is spread over the entire $x$
region. Our results agree with the results of similar studies \cite{buchmann-jpg96}.

\begin{figure}
\includegraphics[width=.9\textwidth] {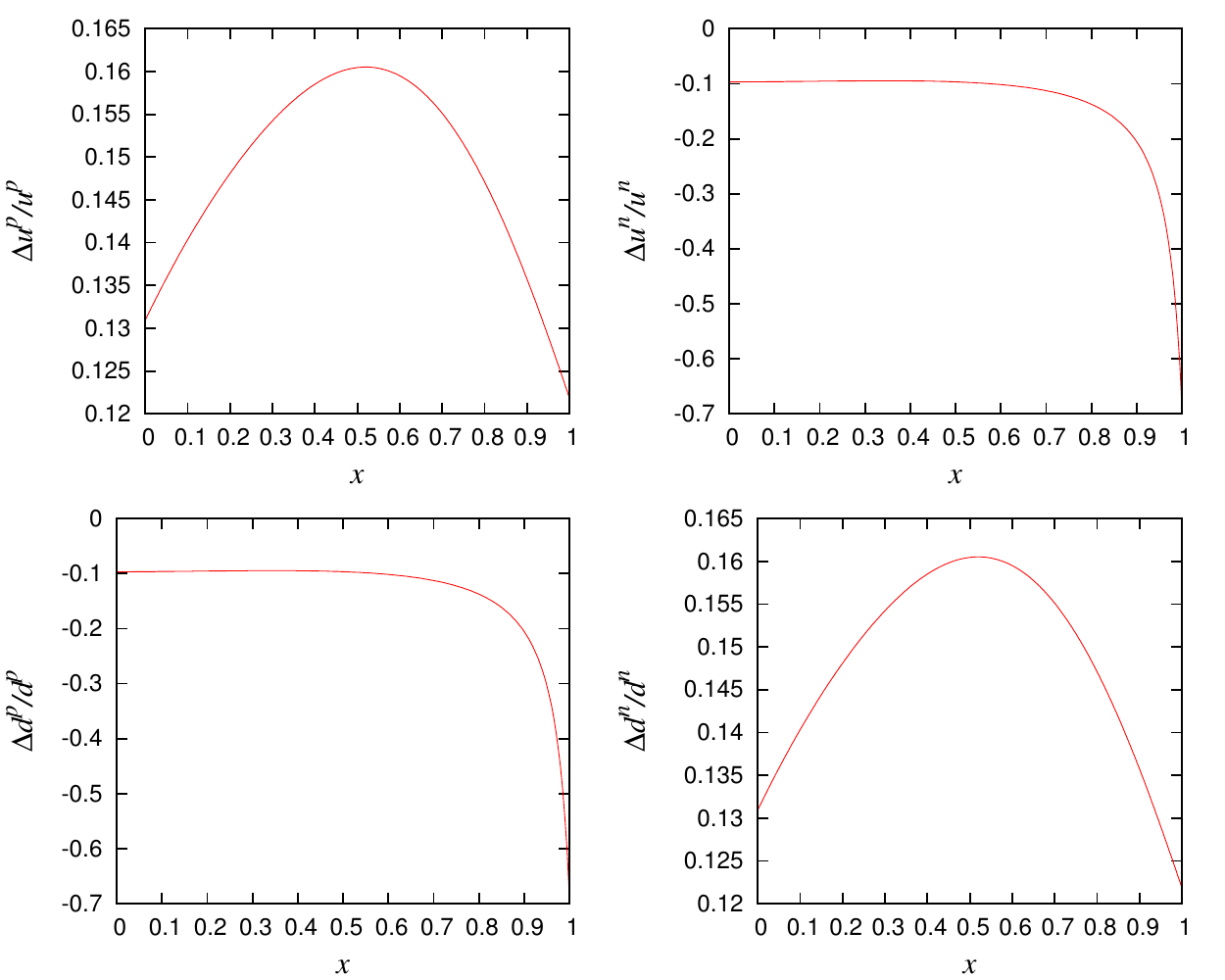}
\caption{(color online). The ratio of polarized to unpolarized distribution functions for the $p$ and $n$: $\frac{\Delta u^p(x)}{u^p(x)}$, $\frac{\Delta u^n(x)}{u^n(x)}$, $\frac{\Delta d^p(x)}{d^p(x)}$, $\frac{\Delta d^n(x)}{d^n(x)}$ as a function of $x$.}
\label{fig2}
\end{figure}

In Fig. \ref{fig2}, the ratio of polarized to unpolarized quark distribution functions for up and down quarks in the $p$ and $n$ $\frac{\Delta u^p(x)}{u^p(x)}$, $\frac{\Delta d^p(x)}{d^p(x)}$ and $\frac{\Delta u^n(x)}{u^n(x)}$,  $\frac{\Delta d^n(x)}{d^n(x)}$ have been presented. It is clear from the figure that $\frac{\Delta d^p(x)}{d^p(x)}$ and $\frac{\Delta u^n(x)}{u^n(x)}$ show constant values at lower and higher $x$ and then suddenly fall off as $x\rightarrow 1$. This is unlike $\frac{\Delta u^p(x)}{u^p(x)}$ and $\frac{\Delta u^n(x)}{u^n(x)}$. As discussed for Fig. \ref{fig1}, the behavior of the unpolarized distribution functions  of $u^p$ and $d^p$ is similar. They first rise at lower $x$ and then fall with $x \rightarrow 1$. However, the behavior of polarized distribution functions $\Delta u^p$ and $\Delta d^p$ is different. In this case, $\Delta u^p$ falls w.r.t $x$ in the positive direction whereas $\Delta d^p$ rises in the negative axis. In the $\frac{\Delta u^p(x)}{u^p(x)}$ graph, both the quantities in the numerator as well as the denominator are positive and fall with $x$ whereas  in the $\frac{\Delta d^p(x)}{d^p(x)}$ graph the numerator is positive while the denominator is negative and rising. The results agree  with the very recent analysis performed by  the Jefferson Lab Angular Momentum (JAM) collaboration
to produce a new parameterization \cite{jam} and the ratio $\frac{\Delta d^p(x)}{d^p(x)}$ was found to remain negative across all $x$. The NQM has the following predictions for the above mentioned quantities
\bea
\frac{\Delta u^p(x)}{u^p(x)}&=& \frac{2}{3}\,, \nonumber \\
\frac{\Delta d^p(x)}{d^p(x)}&=& -\frac{1}{3}\,, \nonumber \\
\frac{\Delta u^n(x)}{u^n(x)}&=& -\frac{1}{3}\,, \nonumber \\
\frac{\Delta d^n(x)}{d^n(x)}&=& \frac{2}{3}\,.
\eea
Since $\Delta u$ and $\Delta d$ denote the difference between the quarks distributions polarized parallel and antiparallel to the polarized nucleon, the distribution when $x \rightarrow 1$ predicts that the structure functions should be dominated by valence quarks polarized parallel to the spin of the nucleon for the case of $\frac{\Delta u(x)}{u(x)}$ and by valence quarks polarized antiparallel to the spin of the nucleon for the case of $\frac{\Delta d(x)}{d(x)}$. Further, dramatically different behaviors for the $\frac{\Delta d(x)}{d(x)}$ ratio in different approaches allowed for $x\gtrsim 0.3$ highlights the critical need for precise data sensitive to the $d$ quark polarization at large $x$ values.
Inclusion of nonzero orbital angular momentum could play an important role
numerically. Further progress on this problem is expected with new data expected from several
experiments at the 12 GeV energy upgraded Jefferson Lab \cite{JLABupgrade} which aim to measure polarization asymmetries of protons up to $x \sim 0.8$.

\begin{figure}
\includegraphics[width=1.\textwidth] {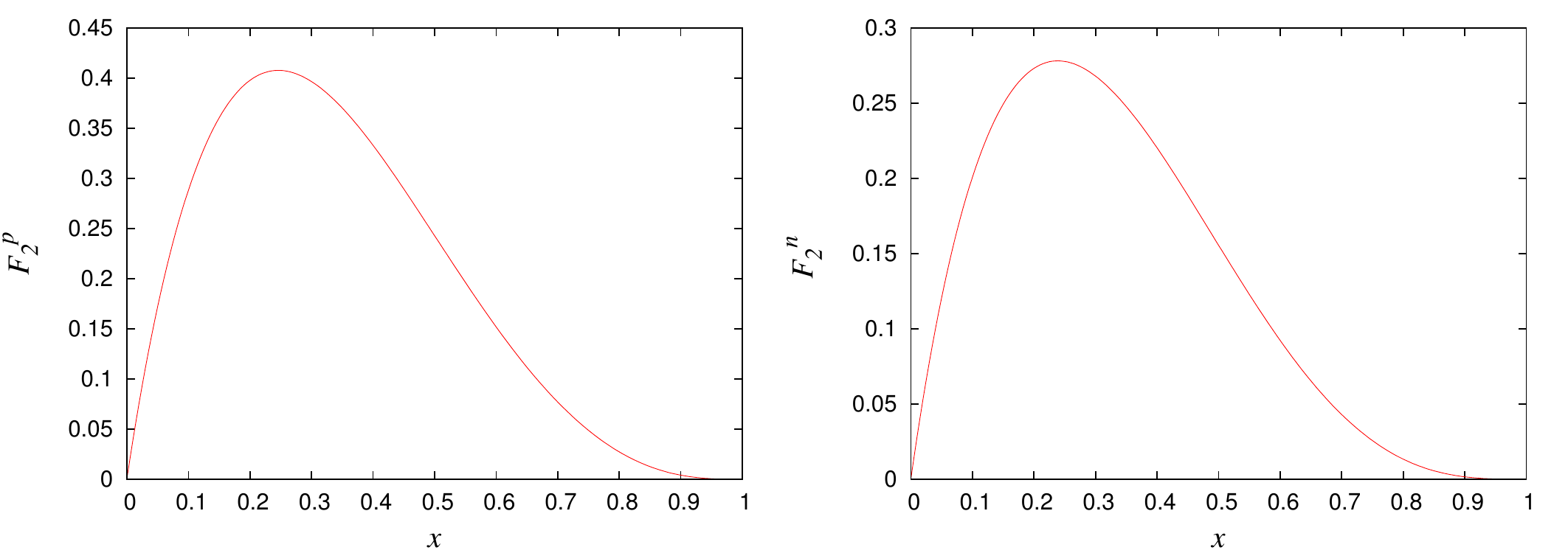}
\caption{(color online). The spin independent structure functions for $p$ and $n$: $F^p_2(x)$ and $F^n_2 (x)$ .}
\label{fig3}
\end{figure}

In Fig. \ref{fig3}, we have plotted the spin independent structure functions $F^p_2(x)$ and $F^n_2(x)$ for the case of $p$ and $n$. The plots clearly project out the distribution of the valence and sea quarks. The function has its peak at around $x\simeq 0.25$. Since the contribution of sea quarks decreases beyond this $x$, the function drops down to zero as $x \rightarrow 1$. There is no mechanism in NQM which can explain the contribution of sea quarks and it has the following predictions for the spin independent structure functions $F^p_1(x)$ and $F^n_1(x)$ at $x \rightarrow 1$
\bea
F^p_1(x)&=& \frac{1}{2}\,, \nonumber \\
F^n_1(x)&=& \frac{1}{6}\,.
\eea
These results may also be related to the Gottfried integral determined from $\frac{F^p_2(x)-F^n_2(x)}{2 x}$. The small $x$ part is suppressed relative to the NQM prediction. As $x \rightarrow 1$, the distribution is dominated by the valence quarks and sea quark asymmetry reduces to zero. This is a clean region to test the valence structure of the nucleon. Measurements of the spin independent structure function in the presently inaccessible low x region will provide crucial information on the low $x$ behavior of $F^p_1(x)$ and $F^n_1(x)$ and also allow access to the non-valence contribution in this region.

\begin{figure}
\includegraphics[width=1.\textwidth] {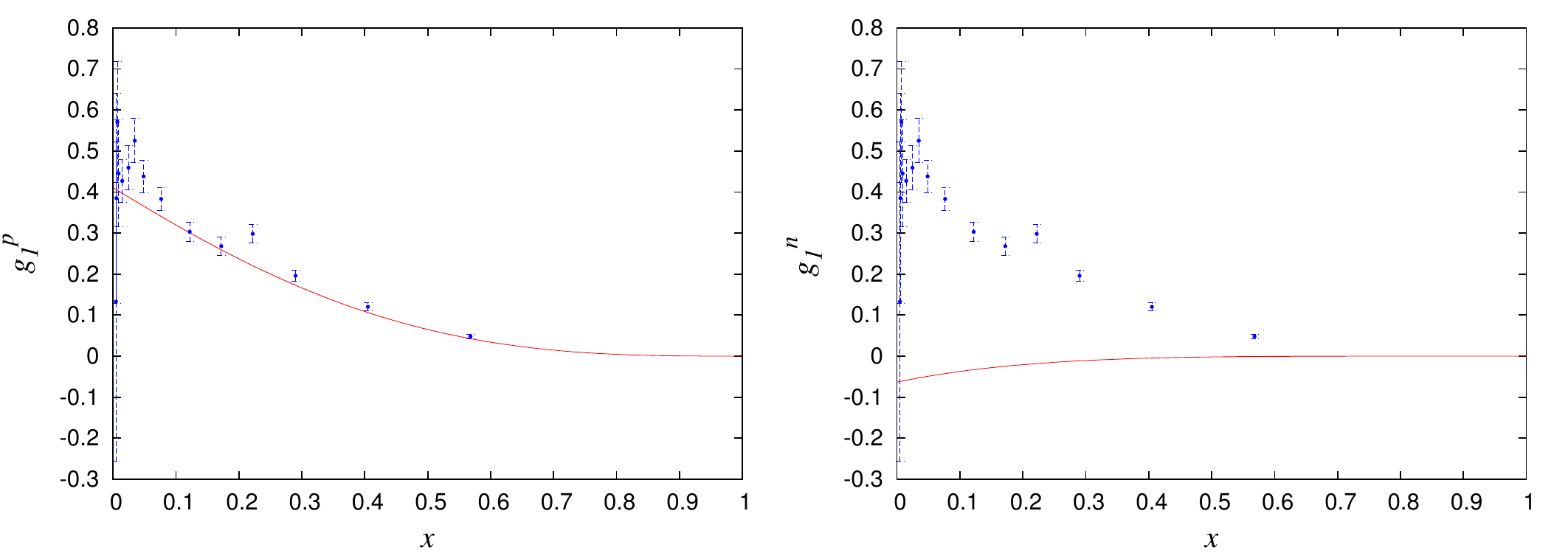}
\caption{(color online). The spin dependent structure functions for $p$ and $n$: $g^p_1(x)$ and $g^n_1 (x)$ compared with the experimental data \cite{smc,hermes_spin,A1p-g1p-Compass-2010}.}
\label{fig4}
\end{figure}

In Fig. \ref{fig4}, we have plotted the spin dependent structure functions $g^p_1(x)$ and $g^n_1(x)$ for the case of $p$ and $n$. For $g^p_1(x)$, we find that it constantly drops down to zero as $x$ increases beyond $x>0.5$   whereas for $g^n_1(x)$, it increases from $-0.07$ to 0 and again at $x>0.5$ it becomes zero.
The NQM predicts
\bea
g^p_1(x)&=& \frac{5}{9}\,, \nonumber \\
g^n_1 (x)&=& 0\,.
\eea It is interesting to note that non-zero values of $g^n_1(x)$ for $x<0.5$ clearly implies the presence of sea quarks. Even though the valence quark distribution is spread over the entire $x$ region and the sea quark distribution decreases with the increasing value of $x$,  the valence and sea quarks are polarized in opposite direction and they mutually cancel the effect of each other at higher values of $x$. When compared with the data \cite{hermes_spin}, we find that our results do not agree with the data at low values of $x$ but as the value of $x$ increases the results are more close.

The results with the data \cite{A1p-g1p-Compass-2010} for the spin dependent structure function $g^p_1(x)$ however agrees to a very large extent both at lower and higher values o $x$. The structure function $g^p_1(x)$ is important also in the context of the measured first moment.  \be
\Gamma_1^p(Q^2)=\int_0^1 g_1^p(x,Q^2) dx=\frac{C_s(Q^2)}{9} g_A^0+\frac{C_{ns}(Q^2)}{12} g_A^3+\frac{C_{ns}(Q^2)}{36} g_A^8.
\ee
It is related to the combinations of the axial-vector coupling constants: $g_A^0$ corresponding to the flavor singlet component, $g_A^3$ and  $g_A^8$ corresponding to the flavor non-singlet components usually obtained from the neutron $\beta-$decay and the semi-leptonic weak decays
of hyperons respectively. Here $C_s$ and $C_{ns}$  are the flavor singlet and non-singlet Wilson coefficients calculable from perturbative QCD. Very recently, a fairly good description of the singlet ($g^0_{A}$) and non-singlet ($g^3_{A}$ and $g^8_{A}$) axial-vector coupling constants has been discussed in the $\chi$CQM \cite{hd}.

\begin{figure}
\includegraphics[width=1.\textwidth] {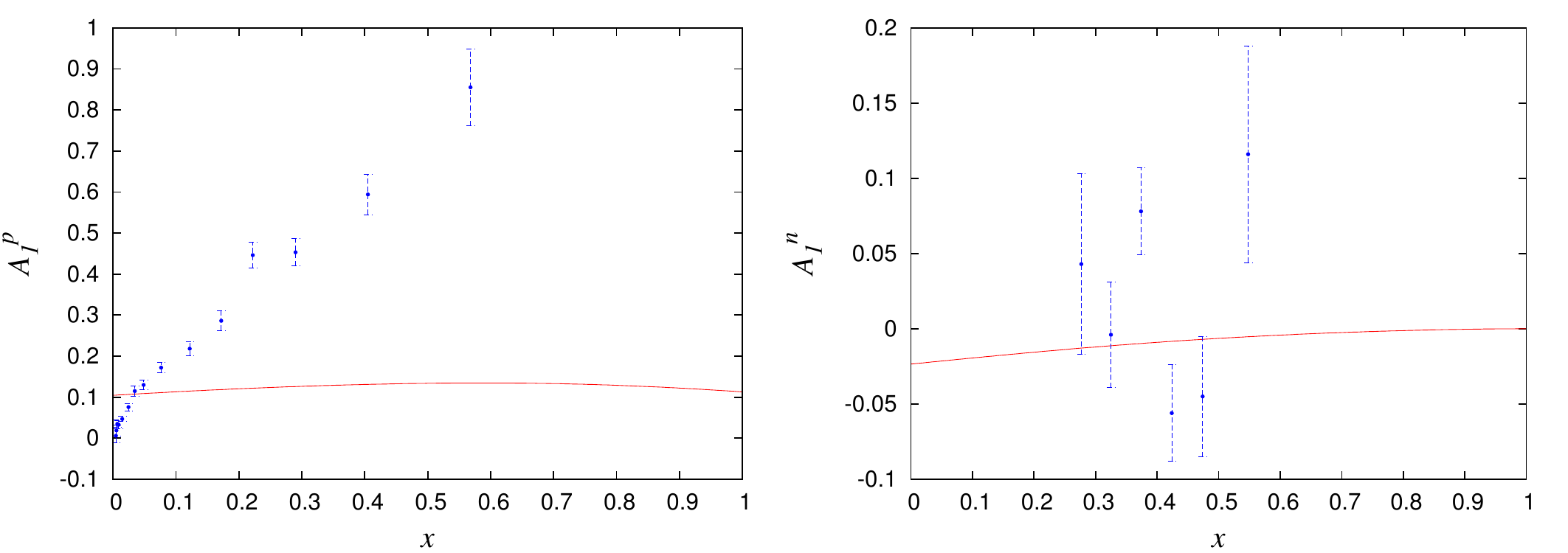}
\caption{(color online). The longitudinal spin asymmetries for $p$ and $n$: $A_1^p(x)$ and $A_1^n(x)$ vs $x$ compared with the experimental data \cite{smc,A1p-g1p-Compass-2010,A1n-Compass-2015}.}
\label{fig5}
\end{figure}

In Fig. \ref{fig5}, the results for $A_1^p(x)$ and $A_1^n(x)$ have been presented. The NQM predictions for these quantities are
\bea
A_1^p(x)&=& \frac{5}{9}\,, \nonumber \\
A_1^n(x)&=& 0\,.
\eea
These results do not agree at all with the experimental results which show that $A_1^p(x)$ increases from 0 at $x \rightarrow 0$ to 1 at $x \rightarrow 1$ \cite{smc,A1p-g1p-Compass-2010}. However, the $A_1^p(x)$ in $\chi$CQM  shows a peak at $x\simeq 0.5$. This low value of  $A_1^p(x)$ at lower and higher values of $x$ be explained on the basis of the sea quarks as in the very low $x$ regime the sea quarks are not highly polarized and at large $x$ there are very few sea quarks and structure is dominated by the valence quarks.
For the case of $A_1^n(x)$, the data \cite{A1n-Compass-2015} is negative at low $x$ and becomes positive at large $x$. In $\chi$CQM, the results agree with the data at some values of $x$ and negative values are obtained. However,  at large $x$, $A_1^n(x)$ continues to remain negative and becomes 0 only at $x \rightarrow 1$. This is because the  $d$ quarks dominate in the valence structure of the $n$ and since they are negatively polarized they keep the values of $A_1^n(x)$ negative. These results agree with the LSS (BBS) parametrization where the Fock states with nonzero quark OAM are included \cite{avakian} where they predict a zero significantly higher $x$.

\begin{figure}
\includegraphics[width=0.55\textwidth] {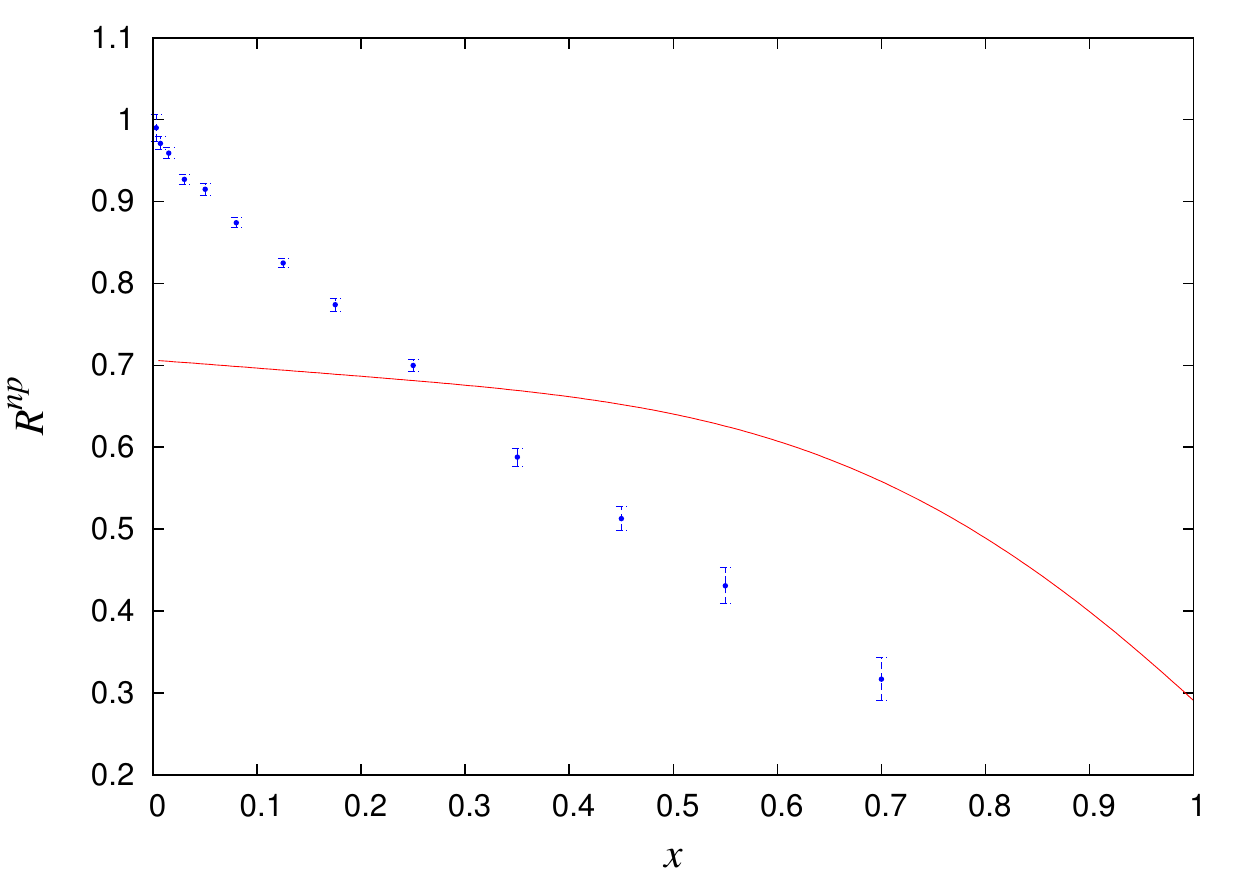}
\caption{(color online). The ratio of the $n$ and $p$ spin independent structure functions $R^{np}(x)=F_2^n(x)/F_2^p(x)$ compared with the experimental data \cite{ratio-nmc}.}
\label{fig6}
\end{figure}

The NQM prediction for the ratio of the neutron and proton structure functions is
\be R^{np} (x)= \frac{2}{3}\,.  \ee
The data \cite{ratio-nmc} however shows that $R^{np}$ drops from 1 at $x \rightarrow 0$ to $\frac{1}{2}$ at $x \rightarrow 1$. From Fig. \ref{fig6}, we find that the $\chi$CQM fits the experimental data quite reasonably. The higher values of $R^{np}$ in the low $x$ region are because of the dominance of the sea quarks in this region. As the value of $x$ increases the valence quarks start dominating leading to the decrease in $R^{np}$.


After having examined the implications of Bjorken scaling variable $x$
for spin independent and spin dependent structure functions,
one would like to study the role of various terms in understanding the
spin sum rule of the nucleon  within the $\chi$CQM.
The results have been presented in Table \ref{total}. The various contributions to the spin sum rule
reveal several interesting points.  In case
the sum rule is to be explained in terms of the spin
polarization contribution of the quarks $S_q$, the orbital
angular momentum of the quarks $L_{q}$ and the total angular momentum of the gluons $J_g$, then these
should add on to give the total spin of the nucleon. The total angular momentum of the gluons cannot be calculated directly
in the present context. It is clear from the results that the valence quark spin and the OAM of the quarks contribute in the same direction to the total proton spin. The sea quark contribution is also significant but in the opposite direction.  Since the $u$ quarks dominate in the valence structure of the proton, the valence spin of the $u$ quarks, quark sea spin of the $u$ quarks and the OAM of the $u$ quarks are higher in magnitude as compared to that of the $d$ quarks. Further, even though the the $u$ quarks carry comparatively larger amount of OAM as compared to the $d$ quarks ($0.265$ as compared to $-0.066$), the total OAM reduces to $0.199$ because of the opposite signs of the $u$ and $d$ quark contributions. The total angular momentum of the $u$ quarks coming from the spin and OAM ($S_u^p+L_u^p$) is 0.777 whereas the total angular momentum of the $d$ quarks ($S_d^p+L_d^p$) is $-0.265$. The contribution of the $s$ quarks to the total angular momentum comes only from the spin part and is $-0.012$. Therefore, the proton spin is dominated by the $u$ quark contribution from the spin as well as OAM. Our results are consistent with the results of Song \cite{song-ijmpa} where it is shown that the quark spin is small, polarization of sea quarks is nonzero and negative and the OAM of sea quarks is parallel to the proton spin. The only difference is that in our model the total angular momentum of the proton has 60\% contribution from the spin of the quarks whereas the OAM contribute 40\% in contrast to the results of Song {\it et al.} \cite{song-ijmpa} where they have around 40\% contribution from the spin of the quarks and 60\% from the OAM.  Our results agree with the model calculations including two-body axial exchange currents necessary to satisfy partial conservation of axial current (PCAC) condition \cite{buchmann-epja-06} as well as with the calculations using spin flavor symmetry based parametrization of QCD \cite{buchmann11}.
It has been shown that the missing spin should be accounted for by the orbital angular momentum of the quarks and antiquarks \cite{myhrer08,thomas09} and the exploration of the angular momentum carried by the quarks and antiquarks is a major aim of the scientific program associated
with the 12 GeV Upgrade at Jefferson Lab \cite{JLABupgrade}. Recently, in a very interesting work, a qualitative interpretation
of the positive and large $u$ quark and small $d$ quark orbital
angular momenta in the proton has been suggested in terms of a prolate
quark distribution corresponding to a positive intrinsic
quadrupole moment \cite{buchmann14}.

\begin{table}
\begin{center}
\begin{tabular}{|c|c|c|c|}       \hline \hline
Parameter  &Data & NQM & $\chi$QM     \\ \hline
$\Delta u^{p}_{\rm V}$ &$-$&$1.333$&$1.333$ \\
$\Delta d^{p}_{\rm V}$ &$-$&$-0.333$&$-0.333$ \\
$\Delta s^{p}_{\rm V}$ &$-$& 0& 0 \\


$\Delta u^{p}_{\rm S}$ &$-$&$0$&$-0.309$ \\
$\Delta d^{p}_{\rm S}$ &$-$&0& $-0.065$ \\
$\Delta s^{p}_{\rm S}$ &$-$&0& $-0.023$ \\
$\Delta u^{p}=\Delta u^{p}_{\rm V}+\Delta u^{p}_{\rm S}$  &$0.85 \pm 0.05$ \cite{PDG} & $1.333$&$1.024$ \\
$\Delta d^{p}=\Delta d^{p}_{\rm V}+\Delta d^{p}_{\rm S}$  & $-0.41 \pm 0.05$ \cite{PDG}& $-0.333$& $-0.398$ \\
$\Delta s^{p}=\Delta s^{p}_{\rm V}+\Delta s^{p}_{\rm S}$  &$-0.07 \pm 0.05$ \cite{PDG}&0& $-0.023$ \\ \hline

$S^p_{q\rm V}=\frac{1}{2} \left (\Delta u^{p}_{\rm V}+\Delta d^{p}_{\rm V}+\Delta s^{p}_{\rm V} \right)$ & $$-$$ &0 & 0.5 \\
$S^p_{q\rm S}=\frac{1}{2} \left (\Delta u^{p}_{\rm S}+\Delta d^{p}_{\rm S}+\Delta s^{p}_{\rm S} \right)$& $$-$$ &0 & $-0.199$ \\
$S^{p}_q=S^{p}_{q\rm V}+S^{p}_{q\rm S}$ &0.30$\pm0.05$ \cite{adams} & 0.5  & 0.301   \\ \hline

$L^p_{u}$ &$-$&$-$& $0.265$ \\
$L^p_{d}$ &$-$&$-$& $-0.066$ \\
$L^p_{s}$ &$-$&$-$& $0$ \\ \hline
$L^p_{q}=L^p_{u}+L^p_{d}+L^p_{s}$ & $$-$$ &0 & 0.199   \\ \hline


$J^{p}_g$ & $$-$$ &0 & 0   \\ \hline \hline
$S^{p}_q+ L^{p}_{q}+J^{p}_g$ & 0.5 & 0.5 & 0.5   \\
\hline \hline
\end{tabular}
\end{center}
\caption{The contributions of various terms to the spin sum rule.} \label{total}
\end{table}

\section{summary and conclusions}

To summarize, the  unpolarized distribution functions of the quarks $q$ and the polarized distribution functions of the quarks $\Delta q$ have been determined  phenomenologically in the chiral constituent quark model ($\chi$CQM). The  $\chi$CQM helps in the understanding the dynamics of the constituents of the nucleon in terms of the explicit contributions of the valence and the sea quarks specifically for the quantities affected by chiral symmetry breaking.
These quantities have important implications in the nonperturbative regime of QCD. In light of precision data available for the low and moderate $x$ region, we have analysed the dependence of various quantities on the Bjorken scaling variable $x$ by incorporating it phenomenologically as the $x<0.3$ is a relatively clean region to test the quark sea structure. In particular, we have computed the spin independent structure functions $F_1^{p,n}(x)$ and $F_2^{p,n}(x)$ as well as the spin dependent  structure functions $g_1^{p,n}(x)$. The implications of the model have also been studied for the $p$ and $n$ longitudinal spin asymmetries $A_1^p(x)$ and $A_1^n(x)$. These asymmetries come from the difference in cross sections in scattering of a polarized lepton from a polarized proton where the leptons are scattered with the same and unlike helicity as that of the proton and one measures the spin dependent structure function $g_1$ via the longitudinal spin asymmetry. Further, the calculations have been extended to compute the explicit ratio of the polarized to un polarized quark distribution functions for up and down quarks in the $p$ and $n$ $\frac{\Delta u^p(x)}{u^p(x)}$, $\frac{\Delta d^p(x)}{d^p(x)}$, $\frac{\Delta u^n(x)}{u^n(x)}$, and $\frac{\Delta d^n(x)}{d^n(x)}$. The $u$ and $d$ quarks have different polarizations and show interesting behavior owing to the dominance of the valence and sea quarks in the different $x$ regions. The qualitative and quantitative role of sea quarks can be further substantiated by discussing the ratio of the $n$ and $p$ spin independent structure functions $R^{np}(x)=\frac{F_2^n(x)}{F_2^p(x)}$. The results have been compared with the recent available experimental observations and the scarcity of precise data at higher $x$ does not allow to favor one model over other. Therefore, new experiments with extended $x$ range are needed for profound understanding of the nonperturbative properties of QCD.
At present, we do not have any deep understanding of the contribution of orbital angular momentum of quarks and the gluon spin, however theoretical studies do indicate that these contributions may not be negligible even in a more
rigorous model. These results will provide important constraints on the future experiments to describe the explicit role of valence and non-valence degrees of freedom.

\section*{ACKNOWLEDGMENTS}

H. D. would like to thank Department of Science and Technology (Ref No. SB/S2/HEP-004/2013), Government of India, for financial support.


\begin{thebibliography}{99}

\bibitem{point-like}  E.D. Bloom {\it et al.}, Phys. Rev. Lett. {\bf 23}, 930 (1969); M. Breidenbach {\it et al.}, Phys. Rev. Lett. {\bf 23}, 935
(1969).

\bibitem{dgg} A. De Rujula, H. Georgi, and S.L. Glashow, Phys. Rev. D {\bf 12}, 147 (1975).

\bibitem{Isgur} N. Isgur, G. Karl and R. Koniuk, Phys. Rev. Lett. {\bf 41}, 1269 (1978);
N. Isgur and G. Karl, Phys. Rev. D {\bf21}, 3175 (1980); N. Isgur {\it et al.},  Phys. Rev. D {\bf 35}, 1665 (1987); P. Geiger and N. Isgur,  Phys. Rev. D {\bf 55}, 299
(1997); N. Isgur, Phys. Rev. D {\bf 59}, 034013 (1999).


\bibitem{yaouanc}  A. Le Yaouanc, L. Oliver, O. Pene, and J.C. Raynal, Phys. Rev. D {\bf 12}, 2137 (1975); A. Le Yaouanc, L. Oliver, O. Pene and J.C. Raynal, Phys. Rev. D {\bf 15}, 844 (1977).

\bibitem{mgupta} M. Gupta, S.K. Sood, and A.N. Mitra, Phys. Rev. D {\bf 16}, 216 (1977); M. Gupta and A.N. Mitra, Phys. Rev. D {\bf 18}, 1585 (1978); M. Gupta, S.K. Sood, and A.N. Mitra, Phys. Rev. D {\bf 19}, 104 (1979); M. Gupta and N. Kaur,
Phys. Rev. D {\bf 28}, 534 (1983); P.N. Pandit, M.P. Khanna, and M. Gupta,
J. Phys. G {\bf 11}, 683 (1985); M. Gupta, J. Phys. G {\bf 16}, L213 (1990).

\bibitem{emc} J. Ashman {\it et al.} (EMC Collaboration), Phys.
Lett. B {\bf 206}, 364 (1988); J. Ashman {\it et al.} (EMC Collaboration), Nucl. Phys. B {\bf 328}, 1 (1989).

\bibitem{smc} B. Adeva {\it et al.} (SMC Collaboration), Phys. Rev.
D {\bf 58}, 112001 (1998); B. Adeva {\it et al.} (SMC Collaboration), Phys. Rev.
D {\bf 60}, 072004 (1999).

\bibitem{adams} P. Adams {\it et al.}, Phys. Rev. D {\bf 56}, 5330
(1997); P.L. Anthony {\it et al.} (E142 Collaboration), Phys. Rev.
Lett. {\bf 71}, 959 (1993); K. Abe {\it et al.} (E143
Collaboration), Phys. Rev. Lett. {\bf 76}, 587 (1996);  K. Abe
{\it et al.} (E154 Collaboration), Phys. Rev. Lett. {\bf 79}, 26
(1997).

\bibitem{hermes_spin} A. Airapetian {\it et al.} (HERMES Collaboration),
Phys. Rev. D {\bf 71}, 012003 (2005); A. Airapetian {\it et al.} (HERMES Collaboration),
Phys. Rev. D {\bf 75}, 012007 (2007).

\bibitem{rev_spin} C.A. Aidala {\it et al.}, Rev. Mod. Phys. {\bf 85}, 655 (2013).

\bibitem{a1np} X. Zheng {\it  et al.}, Phys. Rev. C {\bf 70}, 065207 (2004); D.S. Parno  {\it et al.}, Phys. Lett. B {\bf 744}, 309 (2015).

\bibitem{nmc}  P. Amaudruz {\it et al.} (New Muon Collaboration),
 Phys. Rev. Lett. {\bf 66}, 2712 (1991); M. Arneodo {\it et al.} (New Muon Collaboration),
 Phys. Rev. {\bf D 50}, R1 (1994).

\bibitem{e866} E.A. Hawker {\it et al.} (E866/NuSea Collaboration),
 Phys. Rev. Lett. {\bf 80}, 3715 (1998); J.C. Peng {\it et al.} (E866/NuSea Collaboration),
 Phys. Rev. {\bf D 58}, 092004 (1998); R. S. Towell {\it et al.} (E866/NuSea Collaboration),
{\it ibid.} {\bf 64}, 052002 (2001).

\bibitem{baldit}  A. Baldit {\it et al.} (NA51 Collaboration),  Phys. Lett. B
{\bf 253}, 252 (1994).

\bibitem{hermes_flavor} K. Ackerstaff  {\it et al.} (HERMES Collaboration), Phys. Rev. Lett.
{\bf 81}, 5519 (1998).

\bibitem{gsr} K. Gottfried, Phys. Rev. Lett. {\bf 18}, 1174 (1967).

\bibitem{sample} D.T. Spayde {\it et al.} (SAMPLE Collaboration),
Phys. Lett. B {\bf 583}, 79 (2004).

\bibitem{g0} D. Armstrong {\it et al.} (G0 Collaboration), Phys.
Rev. Lett. {\bf 95}, 092001 (2005). D. Androi$\acute{c}$ {\it et al.} (G0
Collaboration), Phys. Rev. Lett. {\bf 104}, 012001 (2010).

\bibitem{a4} F.E. Maas {\it et al.} (PVA4 Collaboration), Phys. Rev.
Lett. {\bf 93}, 022002 (2004); F.E. Maas {\it et al.} (PVA4 Collaboration), Phys. Rev.
Lett. {\bf 94}, 152001 (2005).

\bibitem{happex} K.A. Aniol {\it et al.} (HAPPEX Collaboration),
Phys. Rev. C {\bf 69}, 065501 (2004); K.A. Aniol {\it et al.} (HAPPEX Collaboration),
Phys. Rev. Lett. {\bf 98}, 032301 (2007); K.A. Aniol {\it et al.} (HAPPEX Collaboration), Eur. Phys. J. A {\bf
31}, 597 (2007); Z. Ahmed {\it et al.} (HAPPEX Collaboration),  Phys. Rev. Lett. {\bf 108}, 102001 (2012),

\bibitem{ji} B.W. Filippone and X. Ji, Adv. Nucl. Phys. {\bf 26}, 1 (2001). X. Ji,
Phys. Rev. Lett. {\bf 78}, 610 (1997).

\bibitem{gluon-nonzero} D. de Florian, R. Sassot, M. Stratmann, W. Vogelsang,  Phys. Rev. Lett.
{\bf 113}, 012001  (2014).

\bibitem{neudis} W.M. Alberico, S.M. Bilenky, and C. Maieron, Phys. Rept. {\bf 358}, 227 (2002) ; U. Dore, Eur. Phys. J. H {\bf 37}, 115 (2012).

\bibitem{cdhs} H. Abramowicz, J.G.H. de Groot, J. Knobloch, J. May, P. Palazzi, A. Para, F. Ranjard, and
J. Rothberg {\it et al.}, Z. Phys. C {bf 15}, 19 (1982); H. Abramowicz {\it et al.}, Z. Phys. C {\bf 17}, 283 (1983); Costa {\it et al.}, Nucl. Phys. B {\bf 297}, 244 (1988).

\bibitem{ccfr1} S.A. Rabinowitz, C. Arroyo, K.T. Bachmann, A.O. Bazarko, T. Bolton, C. Foudas, B. J. King,
and W. Lefmann {\it et al.}, Phys. Rev. Lett. {bf 70}, 134 (1993).

\bibitem{ccfr2} A.O. Bazarko {\it et al.} (CCFR Collaboration and
NuTeV Collaboration), Z. Phys C {\bf 65}, 189 (1995).

\bibitem{charmii} P. Vilain {\it et al.} (CHARM II Collaboration), Eur. Phys. J. C {bf 11}, 19 (1999).

\bibitem{nomad1} P. Astier {\it et al.} (NOMAD Collaboration), Phys. Lett. B {\bf 486}, 35 (2000).
\bibitem{nomad2} O. Samoylov {\it et al.} (NOMAD Collaboration), Nucl. Phys. B {\bf 876}, 339 (2013).
\bibitem{nutev} M. Goncharov {\it et al.} (NuTeV Collaboration), Phys. Rev. D {\bf 64}, 112006 (2001); G.P. Zeller {\it et al.}, Phys. Rev. Lett. {\bf 88}, 091802 (2002); G.P. Zeller {\it et al.}, Phys. Rev. D {\bf 65}, 111103 (2002); D. Mason {\it et al.}, Phys. Rev. Lett. {\bf 99}, 192001 (2007).

\bibitem{chorus} A. Kayis-Topaksu {\it et al.} (CHORUS Collaboration), Nucl. Phys. B {\bf 798}, 1 (2008); A. Kayis-Topaksu {\it et al.}, New J. Phys. {\bf 13}, 093002 (2011).

\bibitem{fermilab} Fermilab E906 proposal, Spokespersons: D. Geesaman and P. Reimer.

\bibitem{jparc} J-PARC P04 proposal, Spokespersons: J.C. Peng and S. Sawada.

%
%
%
%
%
%
%

\bibitem{ellis-brodsky} S.J. Brodsky, J.R. Ellis, and M. Karliner, Phys. Lett. B {\bf 206}, 309 (1988).

\bibitem{alkofer} R. Alkofer, H. Reinhardt, and H. Weigel, Phys. Rept. {\bf 265}, 139 (1996).

\bibitem{christov} K. Goeke, C.V. Christov, and A. Blotz, Prog. Part. Nucl. Phys. {\bf 36}, 207 (1996);
 C.V. Christov, A. Blotz, H.-C. Kim, P. Pobylitsa, T. Watabe, T. Meissner, E. Ruiz Arriola, and K. Goeke,  Prog. Part. Nucl. Phys. {\bf 37}, 91 (1996).

\bibitem{diakonov} D. Diakonov, V.Yu. Petrov, P.V. Pobylitsa, M.V. Polyakov, and C. Weiss, Phys. Rev. D {\bf 56}, 4069 (1997);
D. Diakonov, V.Yu. Petrov, P.V. Pobylitsa, M.V. Polyakov, and C. Weiss,  Phys. Rev. D {\bf 58}, 038502 (1998).



\bibitem{mesoncloud} M. Alberg, E.M. Henley, and G.A. Miller, Phys.Lett. B {\bf 471}, 396 (2000); S. Kumano and M. Miyama, Phys. Rev. D {\bf 65}, 034012 (2002); F.-G. Cao and A.I. Signal, Phys. Rev. D {\bf 68}, 074002 (2003); F. Huang, R.-G. Xu, and B.-Q. Ma, Phys. Lett. B {\bf 602}, 67 (2004); B. Pasquini and S. Boffi, Nucl. Phys. A {\bf 782}, 86 (2007).


\bibitem{wakamatsu} M. Wakamatsu, Phys. Rev. D {\bf 44}, R2631 (1991); M. Wakamatsu, Phys. Rev. D {\bf 46}, 3762 (1992); H. Weigel, L. Gamberg, and H. Reinhardt, Phys. Rev. D {\bf 55}, 6910 (1997); M. Wakamatsu and T. Kubota, Phys. Rev. D {\bf 57}, 5755 (1998); M. Wakamatsu, Phys. Rev. D {\bf 67}, 034005 (2003).

\bibitem{eccm} Y. Ding, R.-G. Xu, and B.-Q. Ma, Phys. Rev. D {\bf 71}, 094014 (2005); L. Shao, Y.-J. Zhang, and B.-Q. Ma, Phys. Lett. B {\bf 686}, 136 (2010).

\bibitem{stat} C. Bourrely, J. Soffer, F. Buccella,  Eur. J. Phys. C {\bf 23}, 487 (2002); I.C. Cl$\ddot{o}$et, W. Bentz, A.W. Thomas,  Phys. Lett. {\bf B621}, 246 (2005); L.A. Trevisan, C. Mirez, T. Frederico, and L. Tomio, Eur. Phys. J. C {\bf 56}, 221 (2008); Y. Zhang, L. Shao, and B.-Q. Ma, Phys. Lett. B {\bf 671}, 30 (2009); Y. Zhang, L. Shao, and B.-Q. Ma, Nucl. Phys. A {\bf 828}, 390 (2009); C.D. Roberts, R.J. Holt, S.M. Schmidt,  Phys. Lett. {\bf B727}, 249 (2013).

\bibitem{bag} A.I. Signal and A.W. Thomas, Phys. Rev. D {\bf 40}, 2832 (1989).

\bibitem{alwall} J. Alwall and G. Ingelman, Phys. Rev. D {\bf 71}, 094015 (2005).

\bibitem{reya} M. Gluck, E. Reya, and A. Vogt, Z. Phys. C {\bf 67}, 433 (1995); M. Gl{\rm $\ddot{u}$}ck, E. Reya, M. Stratmann, and W. Vogelsang, Phys. Rev. D {\bf 53}, 4775 (1996); D. de Florian, C.A. Garcia Canal, and R. Sassot, Nucl. Phys. B {\bf 470}, 195 (1996).

\bibitem{chang-14} J.-C. Peng, W.-C. Chang, H.-Y. Cheng, T.-J. Hou, K.-F. Liu, J.-W. Qiu, Phys. Lett. B {\bf 736}, 411 (2014); W.-C.
Chang, J.-C. Peng, Prog. Part. Nucl. Phys. {\bf 79}, 95  (2014).

\bibitem{manohar} S. Weinberg, Physica A {\bf 96}, 327 (1979); A.
Manohar and H. Georgi, Nucl. Phys. B {\bf 234}, 189 (1984).

\bibitem{eichten} E.J. Eichten, I. Hinchliffe, and C. Quigg, Phys. Rev. {\bf D 45}, 2269
(1992).

\bibitem{cheng} T.P. Cheng and L.F. Li, Phys. Rev. Lett. {\bf
74}, 2872 (1995); Phys. Rev. D {\bf 57}, 344 (1998); Phys. Rev.
Lett. {\bf  80}, 2789 (1998).

\bibitem{johan} J. Linde, T. Ohlsson, and H. Snellman, Phys. Rev. D
{\bf 57}, 452 (1998); {\bf 57}, 5916 (1998).

\bibitem{song} X. Song, J.S. McCarthy, and H.J. Weber, Phys. Rev.
D {\bf 55}, 2624 (1997); X. Song, Phys. Rev. D {\bf 57}, 4114
(1998).

\bibitem{hd} H. Dahiya and M. Gupta, Phys. Rev. D {\bf 64}, 014013
(2001); H. Dahiya and M. Gupta, Phys. Rev. D {\bf 67},
074001 (2003);  H. Dahiya and M. Gupta, Int. Jol.
of Mod. Phys. A, Vol. 19, No. 29, 5027 (2004); H. Dahiya, M. Gupta
and J.M.S. Rana, Int. Jol. of Mod. Phys. A, Vol. 21, No. 21, 4255
(2006); H. Dahiya and M. Gupta,  Phys. Rev. D {\bf 78}, 014001
(2008); N. Sharma, H. Dahiya, P.K. Chatley, and M. Gupta
Phys. Rev. D {\bf 81}, 073001 (2010); N. Sharma and H. Dahiya, Int. Jol. of Mod. Phys. A, Vol. 28, No. 14, 1350052 (2013); H. Dahiya and M. Randhawa, Phys. Rev. D {\bf 90}, 074001 (2014); H. Dahiya, Phys. Rev. D {\bf 91}, 094010 (2015); A. Girdhar, H. Dahiya and M. Randhawa, Phys. Rev. D {\bf 92}, 033012 (2015).

\bibitem{hdmagnetic} H. Dahiya and M. Gupta, Phys. Rev. D
{\bf 66}, 051501(R) (2002); H. Dahiya and M. Gupta, Phys. Rev. D {\bf 67}, 114015 (2003).

\bibitem{nsweak} N. Sharma, H. Dahiya, P.K. Chatley, and M. Gupta,
Phys. Rev. D {\bf 79}, 077503 (2009); N. Sharma, H. Dahiya, and P.K.
Chatley, Eur. Phys. J. A {\bf 44}, 125 (2010).

\bibitem{nres-torres} A.M. Torres, K.P. Khemchandani, N. Sharma, and H. Dahiya, Eur. Phys. Jol. A {\bf 48}, 185 (2012); N. Sharma, A.M. Torres, K.P. Khemchandani, and H. Dahiya, Eur. Phys. Jol. A {\bf 49}, 11 (2013).

\bibitem{charge-radii}	N. Sharma and H. Dahiya,
Pramana, {\bf 81}, 449 (2013);	N. Sharma and H. Dahiya, Pramana, {\bf 80}, 237 (2013).

\bibitem{brodsky} S.J Brodsky, M. Burkardt, and I. Schmidt, Nucl. Phys. B  {\bf 441},
197 (1995); S.J Brodsky, and I. Schmidt, Phys. Lett.  B {\bf 351},
344 (1995).

\bibitem{song-ijmpa} X. Song, Int. Jol. of Mod. Phys. A, Vol. 16, 3673 (2001).

\bibitem{buchmann-jpg96} A. Szczurek, A.J. Buchmann, and A. Faessler,  Jol. of Phys. G {\bf 22}, 1741 (1996).

\bibitem{jam} P. Jimenez-Delgado, A. Accardi, W.Melnitchouk,  Phys. Rev. D {\bf 89}, 034025  (2014); P. Jimenez-Delgado, H. Avakian, W. Melnitchouk Phys. Lett. B {\bf 738}, 263 (2014).

    \bibitem{JLABupgrade} Jefferson Lab experiments PR12-06-109, S. Kuhn {\it et al.}; PR12-06-110, J.-P. Chen {\it et al.}; PR12-06-122, B. Wojtsekhowski {\it et al.}, spokespersons.

\bibitem{A1p-g1p-Compass-2010} M.G. Alekseev {et al.} (COMPASS Collaboration), Phys. Lett. B {\bf 690}, 466 (2010).

\bibitem{A1n-Compass-2015} D.S. Parno {et al.} (Jefferson Lab Hall A Collaboration), Phys. Lett. B {\bf 744}, 309 (2015).

\bibitem{ratio-nmc} P. Amaudruz	{\it et al.}, (New Muon Collaboration),  Nucl. Phys. B {\bf 371}, 3 (1992).

\bibitem{avakian}  H. Avakian, {\it et al.}, Phys. Rev. Lett. {\bf 99},  082001  (2007).

\bibitem{buchmann-epja-06} D. Barquilla-Cano, A.J. Buchmann, and E. Hern$\acute{a}$ndez,  Eur. Phys. Jol. A {\bf 27}, 365 (2006).

\bibitem{buchmann11} A.J. Buchmann and E.M. Henley, Phys, Rev. D {\bf 83}, 096011 (2011).

\bibitem{myhrer08} F. Myhrer and A.W. Thomas, Phys. Lett. B {\bf 663}, 302 (2008).

\bibitem{thomas09} A.W. Thomas, Int. Jol. of Mod. Phys. E {\bf 18}, 1116 (2009).



\bibitem{buchmann14} A.J. Buchmann and E.M. Henley, Few. Body. Sys.  {\bf 55}, 749 (2014).

\bibitem{PDG} K. A. Olive {\it et al.} (Particle Data Group), Chin. Phys. C {\bf 38},
090001 (2014).



\end{thebibliography}
\end{document}